\documentclass[pra,twocolumn]{revtex4-1}
\usepackage{float,graphicx,multirow, amsmath, color, dsfont}
\usepackage{amsmath}
\usepackage{amssymb}
\usepackage{amsfonts}
\usepackage{mathrsfs}
\usepackage{ulem}

\newcommand{\add}[1]{{\color{blue}#1}}
\newcommand{\ket}[1]{\left| #1 \right\rangle}

\begin{document}
\title{Coherence assisted non-Gaussian measurement device
independent quantum key distribution}
\author{Chandan Kumar}
\email{chandankumar@iisermohali.ac.in}
\affiliation{Department of Physical Sciences, Indian
Institute of Science Education and Research (IISER)
Mohali, Sector 81 SAS Nagar, Manauli PO 140306 Punjab India.}
\author{Jaskaran Singh}
\email{jaskaransinghnirankari@iisermohali.ac.in}
\affiliation{Department of Physical Sciences, Indian
Institute of Science Education and Research (IISER)
Mohali, Sector 81 SAS Nagar, Manauli PO 140306 Punjab India.}
\author{Soumyakanti Bose}
\email{soumyabose@iisermohali.ac.in}
\affiliation{Department of Physical Sciences, Indian
Institute of Science Education and Research (IISER)
Mohali, Sector 81 SAS Nagar, Manauli PO 140306 Punjab India.}
\author{Arvind}
\email{arvind@iisermohali.ac.in}
\affiliation{Department of Physical Sciences, Indian
Institute of Science Education and Research (IISER)
Mohali, Sector 81 SAS Nagar, Manauli PO 140306 Punjab India.}
\begin{abstract}
Non-Gaussian operations on two mode squeezed vacuum states
(TMSV) in continuous variable measurement device independent
quantum key distribution (CV-MDI-QKD) protocols have been
shown to effectively increase the total transmission
distances drastically. In this paper we show that photon
subtraction on a two mode squeezed coherent (PSTMSC) state
can further improve the transmission distances remarkably.
To that end we also provide a generalized covariance matrix
corresponding to PSTMSC, which has not been attempted
before. We show that coherence, defined as the amount of
displacement of vacuum state, along with non-Gaussianity can
help improve the performance of prevalent CV-MDI-QKD
protocols. Furthermore, since we use realistic parameters,
our technique is experimentally feasible and can be readily
implemented.
\end{abstract}
\maketitle
%%%%%%%%%%%%%%%%%%%%%%%%%%%%%%%%%%
\section{Introduction}
Quantum key distribution(QKD) protocols~\cite{quantum_crypt, arxiv_qkd_review}
provide a way to carry out unconditionally secure
communication which is not possible in the classical world.
Prevalent QKD schemes can be divided into two main
categories: discrete variable QKD
(DV-QKD)~\cite{ekert_e91,bb84} and continuous variable QKD
(CV-QKD)~\cite{cerf_cvqkd,grosshans_cvqkd,hillery_cvqkd,ralph_cvqkd}.
While the DV-QKD protocols were developed first, CV-QKD are
more readily compatible with current communication
technologies and do not require costly single photon sources
or detectors.  Protocols based on CV systems have also been
shown to be unconditionally secure against collective
attacks~\cite{anthony_sec_proof_cvqkd,
anthony_sec_proof_cvqkd2, anthony_sec_proof_cvqkd3,
renner_sec_proof_cvqkd, shor_sec_proof_bb84} in the finite
key size and asymptotic regime and have been experimentally
implemented~\cite{bennett_exp_qkd,grosshans_exp_cvqkd,qkd_ground_satellite}.

One of the main drawbacks of both the schemes is that in
practice the devices being used may themselves be imperfect
which may lead to serious potential security vulnerabilities
not modelled theoretically. In order to identify all such
security loopholes or side-channels, it is necessary to
fully characterize the devices being used, which in itself
is an arduous task. However, by clever use
of entanglement swapping, it is possible to bypass this
strict characterization of devices which led to the 
development of discrete variable measurement device
independent quantum key distribution(DV-MDI-QKD)
scheme~\cite{braunstein_dvmdiqkd,lo_dvmdiqkd}. In this
scheme a third untrusted party performs
Bell state measurements, whose results are publicly
communicated and used in the process of sharing
the secure key. These protocols have been extensively
analyzed both
theoretically~\cite{curty_dvmdiqkd_theory,ottaviani_dvmdiqkd_theory,wang_dvmdiqkd_theory}
as well as
experimentally~\cite{dcmdiqkd_exp,dvmdiqkd_exp2,dvmdiqkd_exp3}.
Quite soon CV versions of MDI-QKD
were proposed based on
similar ideas of entanglement
swapping~\cite{pirandola_cvmdiqkd_exp,li_cvmdiqkd,ma_cvmdiqkd}.  These protocols
have also been studied
theoretically~\cite{papanastasiou_cvmdiqkd_theory,lupo_cvmdiqkd_theory,wang_cvmdiqkd_self_referenced} 
and realized
experimentally~\cite{pirandola_cvmdiqkd_exp}.  However, the
maximal transmission distances were found to be
unsatisfactory as compared to DV-MDI-QKD. Several
investigations have found that non-Gaussian
operations, like photon addition and subtraction can be used
to increase the entanglement content of the underlying two
mode squeezed
states~\cite{alexei_photon_sub_entanglement,lee_photon_sub_entanglement,zhang_photon_sub_entanglement,bose_nongaussian_entanglement,
dellanno_nongaussian_entanglement,hu_catalysis_entanglement,navarette_photon_sub_entanglement,yang_nongaussian_entanglement}.
It is therefore natural to assume that they can be helpful in improving the maximum
transmission
distances~\cite{ma_cvmdiqkd_photon_sub,zhao_cvmdiqkd_virtual_photon_sub} of CV-MDI-QKD.
Specifically, photon subtraction on two mode squeezed vacuum
state (PSTMSV) has been shown to increase transmission
distances~\cite{ma_cvmdiqkd_photon_sub} as compared to two
mode squeezed vacuum (TMSV) or two mode squeezed
coherent (TMSC) state.  There have also been
indications that coherence might be useful in
improving the efficiency of CV-MDI-QKD protocols, while most
of the analysis has been restricted to non-Gaussian
operations on TMSV states. However, it should also be noted
that application of non-Gaussian operations on the TMSC state is
theoretically quite difficult and a tedious task, which to
the best of our knowledge has not been attempted in full
generality before.

In this paper we show that non-Gaussianity coupled with small
displacements (termed as coherence) of the vacuum state  can
significantly increase the transmission distances, with a
slight decrease in the maximum achievable secure key rate
for the CV-MDI-QKD schemes. We show that transmission
distances can be drastically improved upto $60-70$ Kms using
the same.  Specifically, we apply photon subtraction on the
TMSC state and use it in a general CV-MDI-QKD scheme. We
then show that several previous CV-MDI-QKD protocols based
on either Gaussian or non-Gaussian resources can be
recovered as special cases of our protocol. We provide the
analysis of secure key rate and explicitly identify
coherence as a new and better resource along with
non-Gaussianity for improving transmission distances of
prevalent CV-MDI-QKD for experimentally realizable
parameter range.
In this process we explicitly calculate the
covariance matrix corresponding to $k$-PSTMSC state, which
in itself is quite interesting and can find application in
various other research problems in the field of CV quantum
information processing.

The paper is organized as follows. In
Sec.~\ref{sec:cv_mdi_qkd_pstmsc} we first review the
technique of CV-MDI-QKD and provide a modified version
tailored for PSTMSC state which
is used
throughout the paper. In Sec.~\ref{sec:simulation} we
provide numerical simulations of the performance of PSTMSC
state in CV-MDI-QKD, while in Sec.~\ref{sec:conc} we draw
conclusions from our results and look at future aspects.
%%%%%%%%%%%%%%%%%%%%%%%%%%%%%%%%%
\section{CV-MDI QKD on PSTMSC}
\label{sec:cv_mdi_qkd_pstmsc}
In this section we first review the basic concepts of CV-MDI-QKD 
with a focus on its entanglement based (EB) variant and
then move on to explain our protocol based on photon
subtraction on a two mode squeezed coherent state (PSTMSC). 
 
\subsection{CV-MDI QKD} 
\label{subsec:cv_mdi_qkd} 
In the original entanglement based version of CV-MDI-QKD,
two parties Alice and Bob each prepare a TMSV state with
quadrature variances $V_A$ and $V_B$. 
We assume that $V_A=V_B$ throughout the paper.

The pairs of modes are labelled as $A_1$, $A_2$ and $B_1$,
$B_2$ respectively for Alice and Bob.  Alice and Bob both
transmit one of their modes, $A_2$ and $B_2$, to a third
untrusted party Charlie via quantum channels with lengths
$L_{AC}$ and $L_{BC}$ respectively, while retaining the
modes $A_1$ and $B_1$ with themselves. The total
transmission length is $L=L_{AC}+L_{BC}$.  Charlie
interferes the two modes with the help of a beam splitter
(BS) which has two output modes $C$ and $D$. He then
performs a homodyne measurement of $x$ quadrature on mode
$C$ with outcome $X_C$ and $p$ quadrature on mode $D$ with
outcome $P_D$ and publicly announces the obtained outcomes
$\lbrace X_C, P_D\rbrace$.

With the publicly available knowledge of $\lbrace X_C, P_D
\rbrace$, Bob transforms his retained mode $B_1$ to $B'_1$
by a displacement operation $D(\beta)$, where $\beta= g(X_C+ iP_D)$ and
$g$ is the gain factor. Consequently, the modes $A_1$ and
$B'_1$ become entangled. Later, Alice and Bob
both perform a heterodyne measurement on
the modes $A_1$ and $B'_1$ to obtain the outcomes $\lbrace
X_A, P_A\rbrace$ and $\lbrace X_B, P_B\rbrace$ respectively,
which end up being correlated. Finally both
the parties perform information reconciliation and privacy
amplification to obtain the secret key.
%%%%%%%%%%%%%%%%%%%%%%%%%%%%%%%%%
\subsection{Photon subtraction on two mode squeezed coherent
state}
\label{subsec:photon_subtraction_TMSC}
\begin{figure}
\includegraphics[scale=1]{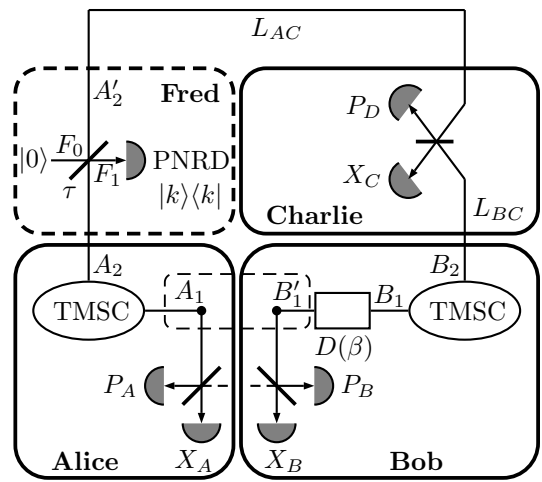}
\caption{Scheme of CV-MDI-QKD using PSTMSC state. The scheme
represents various operations performed by several parties
of which Fred and Charlie are untrusted. After
displacement of mode $B_1$ by Bob based on results announced
by Charlie, the modes $A_1$ and $B'_1$ become correlated.}
\label{fig:photon_sub_protocol}
\end{figure}

In this paper we perform EB CV-MDI-QKD by implementing
photon subtraction on a TMSC state.
The additional parameter in our protocol is the
fact that we are starting with coherent state with a finite
displacement before squeezing it.

Furthermore, we assume
that Bob performs reverse reconciliation (RR), which means
that outcomes obtained by Bob are taken as reference for
Alice to reconcile.  Here, we describe the basic schematic
of the protocol is given in Fig.
\ref{fig:photon_sub_protocol} while relevant calculations
are shown in Appendix with explicit use of phase-space
methods, in particular by using Wigner function description.
We describe the entire protocol through the following steps.

\noindent
\underline{\bf Step 1.} Alice prepares a TMSC state
$|\psi\rangle_{A_1A_2}$ with variance $V_A = \cosh(2 r)$ 
by sending coherent light sources through a non-linear 
optical down converter~\cite{coherent_downconv}, described as
\begin{equation}
|\psi\rangle_{A_1A_2} = S_{12}(r)D_1(d)D_2(d)|00\rangle,
\label{eq:tmsc}
\end{equation}
where
$S_{12}(r)=\text{exp}[r(\hat{a}^\dagger_{A_1}
\hat{a}^\dagger_{A_2}-\hat{a}_{A_1}\hat{a}_{A_2})]$ is 
the squeezing operator with $r$ as a parameter and $D_i(d) =
\text{exp}[d ( \hat{a}^\dagger_{A_i} - \hat{a}_{A_i} ) ]$ 
is the displacement operator displacing mode $A_i$ only
along the $x$ quadrature with magnitude $d$. 
The corresponding Wigner distribution is given by
$W_{\rm{A_{1}A_{2}}}(\xi_{1},\xi_{2})$ 
(see Appendix~\ref{sec_append:Wigner_k-PSTMSC}), where
$\xi_{i} \in \lbrace x_{i},p_{i} \rbrace$ ($i=1,2$).

\noindent
\underline{\bf Step 2.} Alice transmits the mode $A_2$ to
another untrusted party 
Fred who mixes the mode $A_2$ with $F_0$ which is initialized to the 
vacuum state $|0\rangle$, through a BS with transmittivity $\tau$.
This transforms the input state state as $U^{\rm{BS}}_{A_2 F_0}(\tau):
|\psi\rangle_{A_1 A_2 } |0\rangle_{F_0} \rightarrow
|\Psi\rangle_{A_{1} A_{2}^{'} F_{1}}$, 
where $U^{\rm{BS}}_{A_2 F_0}(\tau)$ stands for the
respective BS transformation.
In the same way, corresponding phase-space Wigner distribution 
changes to
$W_{A_{1}A_{2}^{'}F_{1}}(\xi_{1},\xi_{2},\xi_{3})$ (see
Appendix~\ref{sec_append:Wigner_k-PSTMSC}).
 
Fred then performs a measurement on mode $F_1$ with a photon number 
resolving detector (PNRD) represented by the POVM $\lbrace\Pi,
\mathds{1} -\Pi\rbrace$, where 
$\Pi=|k\rangle\langle k|$ is a projection on the $k$ photon state.
Only when $k$ photons are detected on Fred's side the photon
subtraction on the TMSC state 
is considered successful.
This leads to the $k$-PSTMSC given 
by $|{\tilde{\Psi}}\rangle^{k}_{A_{1}A_{2}^{'}} =
{}_{F_{1}}\langle k|\Psi\rangle_{A_{1} A_{2}^{'} F_{1}}$ and 
the corresponding Wigner function as
(Appendix~\ref{sec_append:Wigner_k-PSTMSC}).
\begin{equation}
\tilde{W}_{A_{1}A_{2}^{'}}^{k}(\xi_{1},\xi_{2}) = (-A)^{k}
W_{A_{1}A_{2}^{'}}^{0}(\xi_{1},
\xi_{2})~  L_{k} \left(
\frac{|\xi_{12}|^{2}}{\nu^{2}(\mu^{2} - \tau\nu^{2})}
\right),
\end{equation}
where $A = \frac{\nu^{2}(1 - \tau)}{\mu^{2} - \tau\nu^{2}}$, 
$\mu = \cosh r$, $\nu = \sinh r$, $\xi_{12} = \nu^{2}\sqrt{\tau}(x_{2} + i p_{2}) - 
\mu\nu (x_{1} - i p_{1}) - \frac{d(\mu - \nu)}{2}$ and
$L_{k}(x)$ is the Laguerre polynomial.
$W_{A_{1}A_{2}^{'}}^{0}(\xi_{1},\xi_{2})$ corresponds to the
Wigner distribution for 
$k=0$ case which signifies quantum
catalysis~\cite{hu_catalysis_entanglement}.

It is to be noted that both
$|{\tilde{\Psi}}\rangle^{k}_{A_{1}A_{2}^{'}}$ 
and $\tilde{W}_{A_{1}A_{2}^{'}}^{k}(\xi_{1},\xi_{2})$ are unnormalized. 
Their normalization is given by the probability of $k$
photon subtraction, which is given as
\begin{align}
P_{PS}^{k} &= \sum_{m}\sum_{l}|{}_{A_{1}}\!{\langle} m|
{}_{A_{2}^{'}}\!\langle l| \tilde{\Psi}\rangle^{k}_{A_{1} A_{2}^{'}}|^2
\nonumber
\\
&= \int\frac{dx_{1}dp_{1}}{4\pi} \int\frac{dx_{2}dp_{2}}{4\pi} \tilde{W}_{A_{1}A_{2}^{'}}^{k}(\xi_{1},\xi_{2}).
\end{align} 
In a straightforward calculation it can be shown that (see Appendix~\ref{sec_append:Prob. k-PSTMSC})
\begin{equation}
P^k_{PS}= \frac{A^{k}}{\mu^{2} - \tau \nu^{2}}
e^{-  \frac{d^{2} (1 - \tau) (\mu + \nu)^{2}}{4(\mu^{2} -
\tau \nu^{2})}} 
L_{k}   \left(  -  \frac{d^{2}(\mu + \nu)^{2}}{4 \nu^{2}
(\mu^{2} - \tau \nu^{2})}  \right).
\label{eq:prob_photon_sub}
\end{equation}

\begin{figure}
\includegraphics[scale=1]{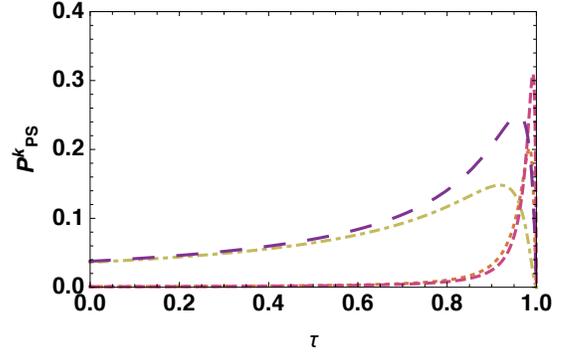}
\caption{Probability of photon subtraction as a function of
BS transmittance $\tau$. Variance is fixed as $V_A=50$ 
and various plots correspond to 1-PSTMSC (Red dashed), 
2-PSTMSC (Orange dotted), 1-PSTMSV (Purple large dashed),
2-PSTMSV (Yellow dash dotted) and TMSV (Black solid).
\label{fig:subtraction}}
\end{figure}

Thus the normalized reduced state for mode $A_1 A'_2$ is
given by 
\begin{align}
|\Psi\rangle^{k}{}_{A_1 A'_2} &=
\left({P^{k}_{PS}}\right)^{-\frac{1}{2}} |{\tilde{\Psi}}\rangle^{k}_{A_{1}A_{2}^{'}}
\nonumber
\\
\rm{or}, W_{A_{1}A_{2}^{'}}^{k}(\xi_{1},\xi_{2}) &= \left({P^{k}_{PS}}\right)^{-1} \tilde{W}_{A_{1}A_{2}^{'}}^{k}(\xi_{1},\xi_{2})
\end{align}
The probability of photon subtraction as a function of $\tau$ for various states is 
shown in Fig.~\ref{fig:subtraction}. It should be noted that the value of $\tau$ used 
throughout the paper is optimized to maximize 
secure transmission length as given in Eqn.~(\ref{eq:secure_keyrate}) and not 
maximizing photon subtraction probability.

We evaluate the corresponding variance matrix in terms of the moment 
generating function defined as (see Appendix~\ref{sec_append:Covariance Matrix k-PSTMSC})
\begin{align}
C_{i,j}^{m,n} &= \left\langle x_{1}^{i}p_{1}^{j}x_{2}^{m}p_{2}^{n} \right\rangle 
\nonumber
\\
&= \int\frac{dx_{1}dp_{1}}{4\pi} \int\frac{dx_{2}dp_{2}}{4\pi} W_{A_{1}A_{2}^{'}}^{k}(\xi_{1},\xi_{2}) x_{1}^{i} p_{1}^{j} x_{2}^{m} p_{2}^{n}.
\end{align} 

We express various moments, e.g., $\langle x_{1}^{l} \rangle$ in a compact notation as $\langle x_{1}^{i} \rangle = C_{i,j}^{m,n} \delta_{i,l}\delta_{j,0}\delta_{m,0}\delta_{n,0}$.
A straightforward calculation yields the variance matrix of the $k$-PSTMSC to be of the following form
\begin{equation}
\Sigma_{A_1 A'_2} = \begin{pmatrix}
V_{A}^{x} & 0 & V_{C}^{x} & 0 \\
0 & V_{A}^{p} & 0 & V_{C}^{p} \\
V_{C}^{x} & 0 & V_{B}^{x} & 0 \\
0 & V_{C}^{p} & 0 & V_{B}^{p}
\end{pmatrix},
\label{eq:variance_a1a2}
\end{equation}
where $V_i^\xi$, $i\in\lbrace A,B,C\rbrace$ and $\xi\in\lbrace
x, p\rbrace$. 
After photon subtraction the mode $A'_2$ is then transmitted
to Charlie via a quantum channel.

\noindent
\underline{\bf Step 3:}
Bob also prepares a TMSC state with
variance $V_B=V_A$ and transmits the mode $B_2$ to Charlie.

\noindent
\underline{\bf Step 4:}
Charlie mixes the two modes 
$A'_2$ $B_2$ received from Alice and Bob respectively
via a BS to obtain output modes $C$ 
and $D$. He then performs a homodyne measurement of
the $x$
quadrature on $C$ and of the $p$ quadrature on $D$. 
The results of these measurements
are declared publicly.

\noindent
\underline{\bf Step 5:}
Based on the publically declared results, Bob adequately displaces his 
retained mode $B_1$ to $B'_1$. Consequently, the two
modes $A_1$ and $B'_1$ are entangled.
Afterwards, both the parties perform heterodyne
measurement on the modes $A_1$ and $B'_1$ respectively and get
correlated outcomes.

\noindent
\underline{\bf Step 6:}
Alice and Bob perform information
reconciliation and privacy amplification to obtain a secure
key string.
%%%%%%%%%%%%%%%%%%%%%%%%%%
\subsection{Special cases}
\label{subsec:spcl}
We now discuss various special cases of PSTMSC state in
the context of CV-MDI-QKD.  In particular, we show that
several of the earlier results in CV-MDI-QKD, can be
obtained as limiting cases of our results on PSTMSC.

In the limit $k\rightarrow 0$, the PSTMSC state simply reduces to the TMSC
state. This we achieve by setting $k=0$ and
$\tau=1$(Fred's BS transmittivity) in our expression for the
covariance matrix. Since the covariance matrix for TMSC is
identical to that of TMSV, the results
obtained for the aforementioned case are identical to the
ones obtained earlier for TMSV. 
Thus, our results with PSTMSC in
CV-MDI-QKD protocols, in the limit $k=0$, reduces to the
earlier results obtained with TMSV~\cite{li_cvmdiqkd}.

Furthermore, in the limit $d \rightarrow 0$ $\forall$ $k \neq
0$, the covariance matrix of PSTMSC represents photon
subtraction on the TMSV state. This specific case has
already been studied extensively in several CV-MDI-QKD
protocols~\cite{subtraction13,ma_cvmdiqkd_photon_sub}. Particularly, the previous 
result on photon subtraction on TMSV is re-examined as 
a special case of ours. It
has also been shown that a non-Gaussian post-selection of
data is equivalent to
PSTMSV~\cite{zhao_cvmdiqkd_virtual_photon_sub,
virtualpra16}. This implies that our results, in the limit
$d\rightarrow 0$, subsume the earlier results on
non-Gaussian CV-MDI-QKD with either photon subtraction on
TMSV or non-Gaussian post-selection. For the rest of the
paper, while quoting and graphically representing results
for states other than PSTMSC, we indeed use the limiting
process described above on our more general state.
%%%%%%%%%%%%%%%%%%%%%%%%%%%%%%%%%%%%%%%%%%%%%%%%%%%
\section{Eavesdropping, channel parameters and secure key
rate}
\label{subsec:eavesdropping}
The protocol described above requires
two quantum channels and one classical channel. We assume
that an eavesdropper, Eve, performs an entangling cloner
attack on each of the quantum channels. The attacks
can be correlated with each other, which is known as a two
mode attack. However, since the two channels
are assumed to be non-interacting and
well separated, the correlated
attack reduces to two independent one-mode collective
attacks on each channel. Under the aforementioned
strategy of Eve, the maximum information gained by her on
the key will be bounded 
by the Holevo bound,
$\chi_{BE}$.  
We note that
the attack considered above is not optimal.

In the following we provide an analysis of various channel
parameters which will be used to calculate the final secret
key. We assume that 
the two channels have transmittance $T_A$ and $T_B$, given as,
\begin{equation}
T_A=10^{-l \frac{L_{AC}}{10}} \quad \text{and} \quad
T_B=10^{-l \frac{L_{BC}}{10}},
\label{eq:transmittance}
\end{equation}
where $l=0.2$dB/Km is the channel loss. Throughout the main
text we consider two cases: 

\noindent
\underline{\bf Symmetric:} In this case we consider
$L_{AC}=L_{BC}$, implying Charlie sits midway between Alice
and Bob. The total transmission
length in this case is $L=2L_{AC}$ with $T_A=T_B$.

\noindent
\underline{\bf Asymmetric:} In this case, $L_{BC}=0$, implying Bob
and Charlie are at the same place. The total transmission
length then becomes
$L=L_{AC}=L_{AB}$ with $T_B=1$.

Bob is the only party who displaces his state and his
outcomes are taken as the reference key to which Alice
reconciles. Due to this inherent asymmetry in the protocol
itself the two cases above give very different results.

We can define a normalized
parameter $T$ associated with channel transmittance in terms
of the only channel parameter $T_A$ as
\begin{equation}
T=\frac{T_A g^2}{2},
\label{eq:T}
\end{equation}
where $g$ is the gain of Bob's displacement operation. The
total channel added noise can then be defined as
\begin{equation}
\chi_{line} = \frac{1-T}{T}+\varepsilon_{th},
\label{eq:chi_line}
\end{equation}
where $\varepsilon_{th}$ is the thermal excess noise in the
equivalent one-way protocol, calculated as given
in~\cite{ma_cvmdiqkd_photon_sub} and written as
\begin{equation}
\varepsilon_{th}=\frac{T_B}{T_A}\left(\varepsilon_B-2\right)+
\varepsilon_A+\frac{2}{T_A},
\label{eq:epsilon_thermal}
\end{equation}
where $\varepsilon_A$ and $\varepsilon_B$ correspond to
thermal excess noise in the respective quantum channels and the 
gain is taken to be as
\begin{equation}
g=\sqrt{\frac{2\left(V_A-1\right)}{T_B\left(V_A+1\right)}},
\label{eq:displacement}
\end{equation}
in order to minimize $\varepsilon_{th}$.

We also assume that Charlie's homodyne detectors are noisy,
with excess noise given as
\begin{equation}
\chi_{homo}=\frac{v_{el}+1-\eta}{\eta},
\label{eq:chi_homo}
\end{equation}
where, $v_{el}$ is the electric noise of the detectors and
$\eta$ is the efficiency. Therefore, the total noise added
because of the 
channel and detectors is
\begin{equation}
\chi_{tot}=\chi_{line}+\frac{2\chi_{homo}}{T_A}.
\label{eq:keyrate}
\end{equation}
The analysis we present in the following sections is for
perfect homodyne detection by Charlie, i.e. $\chi_{homo}=0$,
while in 
Sec.~\ref{subsec:realistic} we explicitly discuss the
tradeoff between Charlie's noise, secure key rate and total
transmission length.

The secure key rate when Eve is assumed to perform a
one-mode collective attack on each quantum channel and under
the aforementioned
channel parameters is given as
\begin{equation}
K=P^k_{PS}\left(\beta I_{AB}-\chi_{BE}\right),
\label{eq:secure_keyrate}
\end{equation}
where $I_{AB}$ is the mutual information between Alice and
Bob and $\chi_{BE}$ 
is the Holevo bound between Bob and Eve, which characterizes
Eve's maximal information on Bob's outcomes. 
It is our intent to 
optimize various parameters for
maximum secure key rate $K$, while
the rest are kept fixed throughout the analysis.

The covariance matrix corresponding to the state $\rho_{A_1
B'_1}$ which is obtained after \textit{Step 5.} of the
protocol given in
Section~\ref{subsec:photon_subtraction_TMSC} is 
\begin{equation}
\Sigma_{A_1 B'_1} = \begin{pmatrix}
V_{A}^{x} & 0 & \sqrt{T}V_{C}^{x} & 0 \\
0 & V_{A}^{p} & 0 & \sqrt{T}V_{C}^{p} \\
\sqrt{T}V_{C}^{x} & 0 & TV'^x_{B} & 0 \\
0 & \sqrt{T}V_{C}^{p} & 0 & TV'^p_{B}
\end{pmatrix},
\label{eq:variance_a1b1}
\end{equation}
where $V'^\xi_B = V^\xi_{B}+\chi_{tot}I_2$ and $V_B^\xi$ is the 
variance of $\xi\in\lbrace x,p\rbrace$
quadrature for Bob's state..
The mutual information between Alice and Bob, $I_{AB}$ can
then be calculated as,
\begin{equation}
I_{AB}=\frac{1}{2}\log_2\left(\frac{V_{A_M}^x}{V_{A_M|B_M}^x}\right)+
\frac{1}{2}\log_2\left(\frac{V_{A_M}^p}{V_{A_M|B_M}^p}\right),
\label{eq:mutualinfo}
\end{equation}
such that,
\begin{equation}
V_{A_M}^\xi = \frac{V_A^\xi+1}{2},
\label{eq:varianceaftermeas}
\end{equation}  
where $V_{A_M|B_M}^\xi$ is the conditional variance of 
Alice's outcome conditioned on Bob's outcome of his
heterodyne measurement and is given as,
\begin{equation}
V_{A_M|B_M}^\xi = \frac{V_{A|B}^\xi+1}{2},
\label{eq:conditionalvarianceaftermeas}
\end{equation}
where, 
\begin{equation}
V_{A|B}^\xi = V_A^\xi-V_C^\xi\left(V_B^\xi+I_2\right)^{-1}(V_C^\xi)^T.
\label{eq:varianceagivenb}
\end{equation}

In order to calculate the upper bound of the information
obtained by Eve, we assume that she 
also has access to Fred's mode $F$ and her state is then
given by $\rho_{EF}$. We also assume that 
she can purify $\rho_{A_1B'_1EF}$. The Holevo bound
$\chi_{BE}$ between Bob and Eve can then be calculated as,
\begin{equation}
\begin{aligned}
\chi_{BE}&=S(\rho_{EF})-\int dm_B p(m_B)S(\rho_{EF}^{m_B})\\
&=S(\rho_{A_1 B'_1}) - S(\rho_{A_1}^{m_{B'_1}}),
\label{eq:holevo}
\end{aligned}
\end{equation}
where $S(\rho)$ is the von-Neumann entropy of the state
$\rho$, $m_B$ represents measurement outcomes of Bob with
probability
density $p(m_B)$ and $\rho_{EF}^{m_B}$ is the state of Eve
conditioned on Bob's outcome. The covariance matrices
corresponding to the 
states $\rho_{A_1B'_1}$ and $\rho_{A_1}^{m_{B'_1}}$ are
represented by $\Sigma_{A_1 B'_1}$ and
$\Sigma_{A_1}^{m_{B'_1}}$ respectively. The 
von-Neumann entropy $S(\rho_{A_1 B'_1})$ and
$S(\rho_{A_1}^{m_{B'_1}})$ are functions of symplectic
eigenvalues $\lambda_1$, $\lambda_2$ of $\Sigma_{A_1 B'_1}$
and $\lambda_3$ of $\Sigma_{A_1}^{m_{B'_1}}$
which are given as,
\begin{equation}
S(\rho_{A_1 B'_1}) = G\left[\frac{\lambda_1-1}{2}\right]+
G\left[\frac{\lambda_2-1}{2}\right],
\label{eq:sab}
\end{equation}
and
\begin{equation}
S(\rho_{A_1}^{m_{B'_1}}) = G\left[\frac{\lambda_3-1}{2}\right],
\label{eq:sa}
\end{equation}
with,
\begin{equation}
G(x)=(x+1)\log_2(x+1)-x\log_2x,
\label{eq:thermalstate}
\end{equation}
is the von-Neumann entropy of the thermal state.
%%%%%%%%%%%%%%%%%%%%%%%%%%%%%%%%%%%%%%

\section{Simulation results}
\label{sec:simulation}
Having described the protocol in its generality, in this
section we provide numerical results corresponding to
optimization of some parameters while the rest are kept fixed.  We
provide the analysis for both the symmetric and asymmetric
case, respectively.

It was shown in~\cite{ma_cvmdiqkd_photon_sub} that photon
subtraction on a TMSV state can lead to an increase in
transmission distances for secure key rate for QKD,
especially in the extreme asymmetric case, where distances
can reach up to $60-70$ kms approximately as compared to
$40-50$ kms achievable using only TMSV. In the following
subsections we show that performing photon subtraction on a
TMSC state can lead to a much better performance \add{as}
compared to TMSV.
%%%%%%%%%%%%%%%%%%%%%%%%%%%%%%%%%%%%%%%%%%%%%
%
\subsection{Effect of displacement on distance for a fixed key rate}
\label{subsec:c_vs_l}
\begin{figure}
\includegraphics[scale=1]{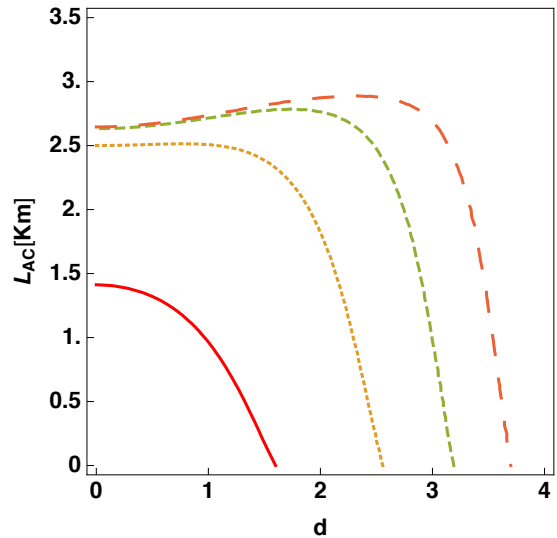}
\caption{
Plots of $L_{AC}$ as a function of vacuum state
displacement $d$ for different values of the secret key for
the symmetric case with one photon subtraction.
The different values of the fixed 
secret key rate are be $K=10^{-1}$ (Red solid),
$K=10^{-2}$ (Tiny dashed), $K=10^{-3}$ (Dashed) and 
$K=10^{-4}$ (large dashed). 
The total transmission length is
$L=2L_{AC}$. The other parameters are 
fixed as $V_A = 50$, $\eta = 1$, $\varepsilon_A^{th} = 0.002= 
\varepsilon_B^{th}$ and $\beta = 96\%$.}
\label{fig:a_vs_length_symm}
\end{figure}
\begin{figure}
\includegraphics[scale=1]{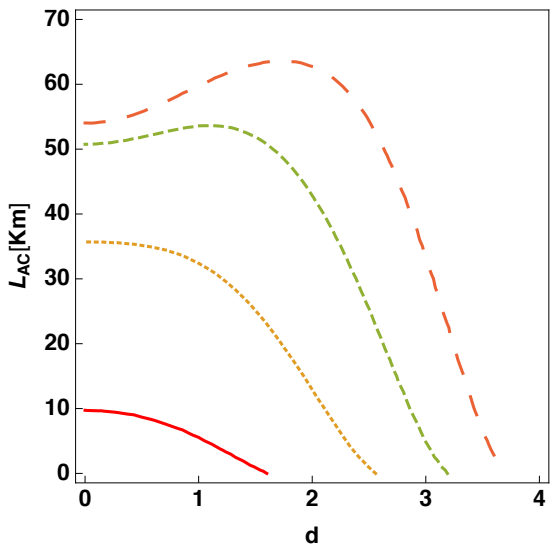}
\caption{
Plots of $L_{AC}$ as a function of vacuum state
displacement $d$ for different values of the fixed secret key for
the extreme asymmetric case with one photon subtraction.
The different values of the fixed secret key rates are $K=10^{-1}$ (Red solid),
$K=10^{-2}$ (Tiny dashed), $K=10^{-3}$ (Dashed) and 
$K=10^{-4}$ (large dashed). The total transmission length is
$L=L_{AC}$ and parameters are 
fixed as $V_A = 50$, $\varepsilon_A^{th} = 0.002= 
\varepsilon_B^{th}$ and $\beta = 96\%$.}
\label{fig:a_vs_length_asymm}
\end{figure}
Photon subtraction on TMSV state is known to increase
transmission distances. We show that
transmission distances can be further improved if we
consider PSTMSC state. We explore the variation of distance
with displacement for symmetric as well extreme asymmetric
case for several fixed values of the key rate. The results
are shown in
Figs.~\ref{fig:a_vs_length_symm} and
\ref{fig:a_vs_length_asymm}.

For the symmetric case, we find that a secure key rate of
$10^{-4}$ bits/pulse can be achieved for $0.5$ kms more than
what was possible for PSTMSV and is shown in
Fig.~\ref{fig:a_vs_length_symm}.  A much more significant
improvement of approximately $10$ kms is possible in the
asymmetric case for the same key rate as shown in
Fig.~\ref{fig:a_vs_length_asymm}.

Both Fig.~\ref{fig:a_vs_length_symm} and
Fig.~\ref{fig:a_vs_length_asymm} imply an increase in
transmission distance with an increase in coherence with a much
more significant difference in the extreme asymmetric case.
We find that setting the coherence, $d=2$, transmission
lengths can be significantly improved. However, it is also
seen that larger coherence values are detrimental to the
protocol.
%%%%%%%%%%%%%%%%%%%%%%%%%%%%%%%%%%%%%%%%%%%%%%%%%
\subsection{Effect of Variance on key rate for a fixed
distance}
\label{subsec:v_vs_k}
\begin{figure}
\includegraphics[scale=1]{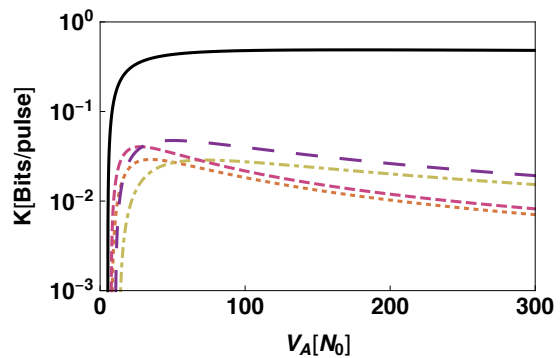}
\caption{Secret key rate as a function of $V_A$ in the
symmetric case where
$L_{AC}=L_{BC} = 2$ kms and total transmission distance $L=2
L_{AC}$. Parameters are 
fixed as: $\tau=0.9$, $\varepsilon_A^{th} = 0.002=
\varepsilon_B^{th}$, $\beta = 96\%$ and displacement $d=2$.
Various curves correspond to $1$-PSTMSC (Red dashed), 
$2$-PSTMSC (Orange dotted), $1$-PSTMSV (Purple large
dashed), $2$-PSTMSV (Yellow dash dotted) and TMSV (Black
solid).}
\label{fig:vvsksymm}
\end{figure}
\begin{figure}
\includegraphics[scale=1]{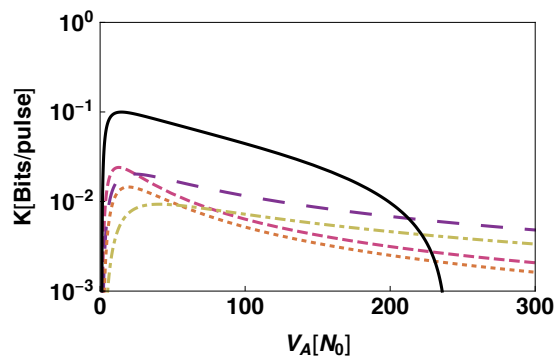}
\caption{Secret key rate as a function of $V_A$ in the
asymmetric case where
$L_{AC} = 20$ kms, $L_{BC}=0$ and total transmission
distance $L=L_{AC}$. Parameters are fixed as: $\tau=0.9$,
$\varepsilon_A^{th} = 0.002=
\varepsilon_B^{th}$, $\beta = 96\%$ and displacement $d=2$.
Various curves correspond to $1$-PSTMSC (Red dashed), 
$2$-PSTMSC (Orange dotted), $1$-PSTMSV (Purple large
dashed), $2$-PSTMSV (Yellow dash dotted) and TMSV (Black
solid).}
\label{fig:vvskasymm}
\end{figure}
Like coherence and optimal transmittivity of photon
subtraction, choice of variance also plays a vital role in
CV-MDI-QKD. In Fig.~\ref{fig:vvsksymm}
and Fig.~\ref{fig:vvskasymm} we plot secure key rate as a
function of variance of Alice's PSTMSC state while keeping
all the other parameters fixed. 
We see that for a fixed transmission length $L_{AB}=4$ kms
in the symmetric case, the TMSV state outperforms all the
other states including PSTMSC. 
It is also seen that PSTMSV state can achieve a higher
secure key rate than PSTMSC for a fixed length and variance.
In the extreme asymmetric case, 
where we fix length $L_{AB}=L_{AC}=20$ kms, the TMSV state
still outperforms the other states in terms of higher secure
key rate for lower variance.
However, for variances larger than $250$, the other states
including PSTMSV and PSTMSC outperform TMSV. It is also seen
that for extremely small value 
of variance, PSTMSC state provides a slightly higher key
rate than PSTMSV. However, we find that a small value of
variance, $V_A=50$, is enough in order 
to optimize the protocol for longer transmission lengths.

\subsection{Effect of Length on  key rate}
\label{subsec:l_vs_k}
\begin{figure}
\includegraphics[scale=1]{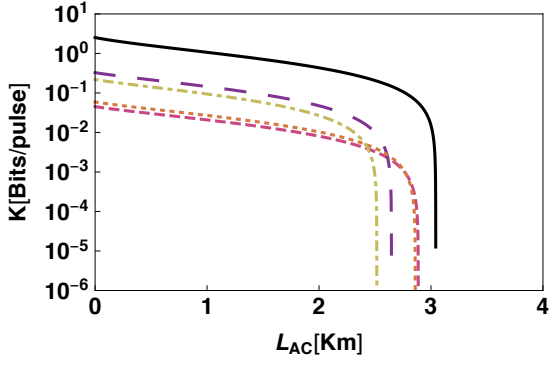}
\caption{Secret key rate as a function of $L_{AC}$ in the
symmetric case where
$L_{AC}=L_{BC}$, $V_A=50$ and total transmission distance
$L=2 L_{AC}$. Parameters are fixed as: $\tau=0.9$,
$\varepsilon_A^{th} = 0.002=
\varepsilon_B^{th}$, $\beta = 96\%$ and coherence $d=2$.
Various curves correspond to $1$-PSTMSC (Red dashed), 
$2$-PSTMSC (Orange dotted), $1$-PSTMSV (Purple large
dashed), $2$-PSTMSV (Yellow dash dotted) and TMSV (Black
solid).}
\label{fig:lvsksymm}
\end{figure}
\begin{figure}
\includegraphics[scale=1]{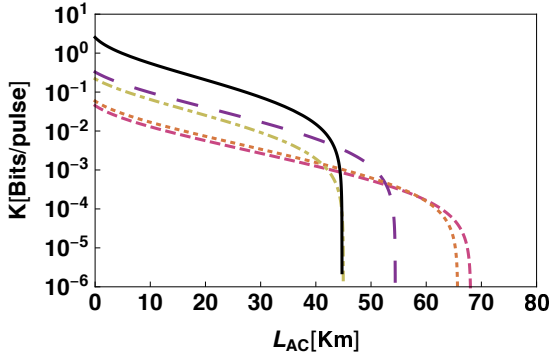}
\caption{Secret key rate as a function of $L_{AC}$ in the
asymmetric case where
$L_{BC}=0$, $V_A=50$ and total transmission distance
$L=L_{AC}$. Parameters are fixed as: $\tau=0.9$,
$\varepsilon_A^{th} = 0.002=
\varepsilon_B^{th}$, $\beta = 96\%$ and displacement $d=2$.
Various curves correspond to $1$-PSTMSC (Red dashed), 
$2$-PSTMSC (Orange dotted), $1$-PSTMSV (Purple large
dashed), $2$-PSTMSV (Yellow dash dotted) and TMSV (Black
solid).}
\label{fig:lvskasymm}
\end{figure}

Equipped with 
approximate values of $\tau$, $a$ and $V_A$ that maximize
transmission distance for PSTMSC, we plot key rate as a
function of $L_{AC}$ in Fig.~\ref{fig:lvsksymm}
and Fig.~\ref{fig:lvskasymm} corresponding to symmetric and
extreme asymmetric case respectively. In the symmetric case
the total transmission distance
is given by $L_{AB}=2L_{AC}$, while in the extreme
asymmetric case, $L_{AB}=L_{AC}$. 

We find that for the symmetric case TMSV state offers better
results than any other state in terms of both key rate and
transmission distance. However, for the same case it is
seen that PSTMSC state offers a longer transmission length
than PSTMSV while the maximum key rate achievable by the
former is less than the latter. 

The advantage of 
using PSTMSC is apparent in the extreme asymmetric case
where it significantly outperforms the others by
approximately $10$ kms while its maximum achievable key rate
is still the least.

From the above it is clear that a small amount of coherence
is actually favorable if maximizing the transmission
distance. Although offering lesser key rate, PSTMSC state
can drastically increase the distances upto which QKD can be
performed. Furthermore, real experiments rarely deal with
squeezed vacuum states which are also harder to prepare than
squeezed coherent states. 
Our results make it evident that
it is highly advantageous if a squeezed coherent light
source is used instead of squeezed vacuum.

\subsection{Noisy homodyne detectors}
\label{subsec:realistic}
\begin{figure}
\includegraphics[scale=1]{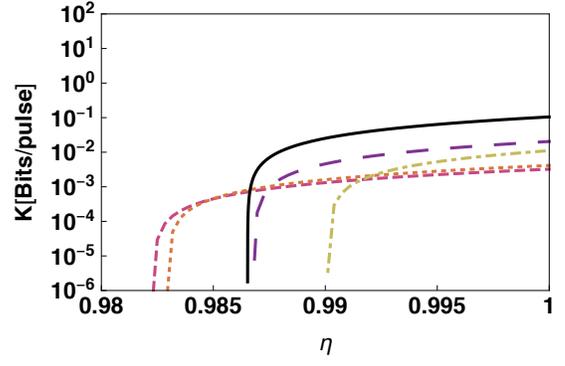}
\caption{Secret key rate as a function of $\eta$ in the
asymmetric case where
$L_{AC} = 20$ kms and $V_A=50$ with total transmission
distance $L=L_{AC}$. Parameters are fixed as: $\tau=0.9$,
$\varepsilon_A^{th} = 0.002=
\varepsilon_B^{th}$, $\beta = 96\%$, displacement $d=2$ and
$\nu_{el}=0.01$. Various curves correspond to $1$-PSTMSC
(Red dashed), 
$2$-PSTMSC (Orange dotted), $1$-PSTMSV (Purple large
dashed), $2$-PSTMSV (Yellow dash dotted) and TMSV (Black
solid).}
\label{fig:eta_vs_k}
\end{figure}
\begin{figure}
\includegraphics[scale=1]{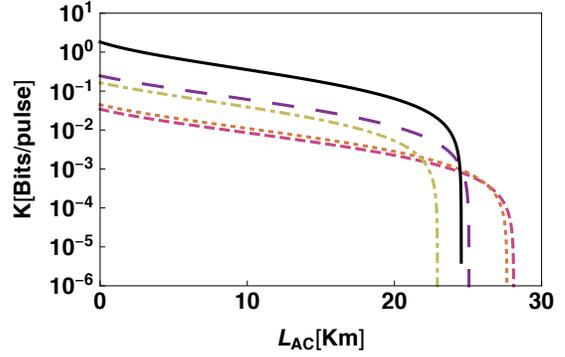}
\caption{Secret key rate as a function of $L_{AC}$ in the
asymmetric case where
$V_A= 50$ and total transmission distance $L=L_{AC}$.
Parameters are fixed as: $\tau=0.9$, $\varepsilon_A^{th} =
0.002=
\varepsilon_B^{th}$, $\beta = 96\%$, displacement $d=2$ and
detector efficiency $\eta =0.995$. Various curves correspond
to $1$-PSTMSC (Red dashed), 
$2$-PSTMSC (Orange dotted), $1$-PSTMSV (Purple large
dashed), $2$-PSTMSV (Yellow dash dotted) and TMSV (Black
solid).}
\label{fig:realistic_graph}
\end{figure}
Noise and efficiency of Charlie's homodyne detection also
plays a major role in optimizing the total transmission
length of the protocol.
We find that the efficiency of the detectors needs to be
close to unity to maintain transmission lengths upto $30$ kms as is
evident in Fig.~\ref{fig:realistic_graph}.
A drastic drop of approximately $30$ kms in the transmission
length is observed for even a small detector inefficiency of
$\eta =0.995$ and $\nu_{el}=0.01$. Furthermore, the range
for tolerable detector efficiency $\eta$ can be made as low
as $86\%$ for smaller transmission distances. From Fig.~\ref{fig:eta_vs_k} it 
is also seen that PSTMSC state is more robust than PSTMSV
and TMSV against detection inefficiencies which is also
evident in Fig.~\ref{fig:realistic_graph}. 

%%%%%%%%%%%%%%%%%%%%%%%%%%%%%%%%
\section{Conclusion}
\label{sec:conc}
In this paper we showed that PSTMSC states have an
advantage over PSTMSV in terms of the distance over which
QKD can be carried out.
To that end we explicitly derive the covariance
matrix for PSTMSC state, which to the best of our knowledge 
has not been attempted before. Using the same in CV-MDI-QKD, we
find that for perfect homodyne detectors, and a small
amount of coherence $d=2$, the transmission
distances can be made as large as $70$ kms in the extreme
asymmetric case which is
considerably
longer than what is achievable with PSTMSV. However, the same effects are not
noticeable in the symmetric case.
We also find that transmission distances of PSTMSC MDI QKD can
achieve a maximum value of $30$ kms under noisy homodyne
detectors of Charlie, which is suitable for a small metropolitan city.
This distance is again significantly higher than what
can be achieved with PSTMSV with noisy homodyne detectors, implying that PSTMSC is a
better candidate for long distance CV-MDI-QKD. Furthermore,
we showed that many of the previous results of CV-MDI-QKD
can be obtained as limiting cases of
the
PSTMSC based CV-MDI-QKD protocol.
We emphasize that, 
while non-Gaussian operations on TMSV have been well
studied, the same is not true for TMSC states. The covariance
matrix computed here is expected to be useful  
to characterize the properties of such states and will find further 
application in various information processing tasks.

\begin{acknowledgements}
J.S. would like to acknowledge funding from UGC, India.
Arvind acknowledges funding from DST India under
Grant No. EMR/2014/000297.
\end{acknowledgements}

\appendix
\section{Wigner Distribution for $k$-PSTMSC state}
\label{sec_append:Wigner_k-PSTMSC}

\begin{figure}[H]
\centering
\includegraphics[scale=1]{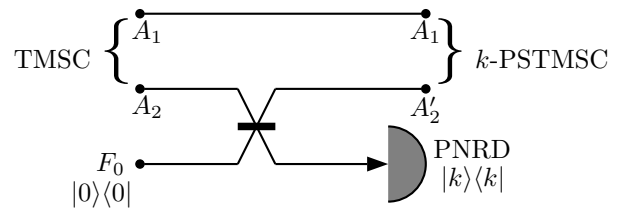}
\caption{A TMSC state is passed through the modes $A_1$
$A_2$ while mode $F_0$ is initialized to 
vacuum $|0\rangle\langle 0|$. 
Photon number resolving detector (PNRD) given by the POVM
$\lbrace \Pi, \mathds{1}-\Pi\rbrace$,
where $\Pi = |n\rangle\langle n|$ is applied to mode $F_0$.}
\label{fig:photon_sub_machine} 
\end{figure}

In this section we provide detailed calculations for evaluating the Wigner distribution for $k$-PSTMSC state.
We start with a TMSC state and a third mode initialized to the vacuum state. After mixing the vacuum state with 
one of the modes of TMSC by the use of a BS, we perform a $k$ photon number detection on the third mode. 
Consequently, the TMSC state is then transformed to a $k$-PSTMSC. The basic schematic of photon subtraction 
on a TMSC is shown in Fig.~\ref{fig:photon_sub_machine}.

We start with the Wigner distribution for TMSC.
In shot noise unit (SNU), for a two-mode squeezed coherent state (TMSC), i.e., $\rho_{A_{1}A_{2}}=S_{ab}(r)\ket{d,d}$, with $S_{ab}(r)=e^{r(a^{\dagger}b^{\dagger}-ab)}$, its Wigner distribution is given by
\begin{widetext}
\begin{equation}
W_{A_{1}A_{2}}(\xi_{1},\xi_{2}) = 4 e^{-d^{2}}  e^{-\frac{\mu^{2} + \nu^{2}}{2}(x_{1}^{2} + p_{1}^{2} + x_{2}^{2} + p_{2}^{2}) + 2\mu\nu(x_{1}x_{2} - p_{1}p_{2}) + d(\mu - \nu)(x_{1} + x_{2})},
\label{append_eq_exp_WigDist_TMSC}
\end{equation}
\end{widetext}
where $\mu=\cosh r$ and $\nu=\sinh r$ and the normalization is given by $ \int\frac{dx_{1}dp_{1}}{4\pi}   \int\frac{dx_{2}dp_{2}}{4\pi} W_{\rm{A_{1}A_{2}}}(\xi_{1},\xi_{2}) = 1$ such that $\xi_{i}=\lbrace x_{i},p_{i} \rbrace$ ($i=1,2$).
On the other hand, Wigner distribution for a single mode photon number state is given by
\begin{align}
W^{n}(\xi)&= 2 (-1)^{n} e^{-\frac{x^{2} + p^{2}}{2}} L_{n} (x^{2} + p^{2})
\nonumber
\\
&= \frac{2 (-1)^{n} e^{-\frac{x^{2} + p^{2}}{2}}}{n!} \partial^{n}_{\eta}\partial^{n}_{\zeta} \left[ e^{\eta\zeta + (x+ip)\eta - (x-ip)\zeta} \right]_{\substack{\eta=0\\ \zeta=0}},
\label{append_eq_exp_WigDist_photon}
\end{align}
where $L_{n}(x)$ is the Laguerre polynomial.  Wigner
Distribution for the vacuum ($W^{0}(\xi)$) could be
trivially obtained by setting $n=0$.

The tripartite Wigner distribution for Alice, Bob in TMSC
and Fred in vacuum is given by
$W_{A_{1}A_{2}F_{0}}(\xi_{1},\xi_{2},\xi_{3}) =
W_{A_{1}A_{2}}(\xi_{1},\xi_{2}) W_{F_{0}}^{0}(\xi_{3})$.
Let's consider the mixing of modes "$A_{2}$" and "$F$" through BS
with transmittivity $\tau$ for which quadrature variables
transform as $\begin{pmatrix} \xi_{2} \\ \xi_{3}
\end{pmatrix} \rightarrow \begin{pmatrix} \xi^{'}_{2} \\
\xi^{'}_{3} \end{pmatrix} = S_{\rm{BS}}\begin{pmatrix}
\xi_{2} \\ \xi_{3} \end{pmatrix}$, where \begin{equation}
S_{\rm{BS}}=\begin{pmatrix} \sqrt{\tau} & 0 & \sqrt{1-\tau}
& 0 \\ 0 & \sqrt{\tau} & 0 & \sqrt{1-\tau} \\ -\sqrt{1-\tau}
& 0 & \sqrt{\tau} & 0 \\ 0 & -\sqrt{1-\tau} & 0 &
\sqrt{\tau} \end{pmatrix} \label{append_eq_exp_mat_bs}
\end{equation} is the BS transformation matrix.  It is
well-known that under a linear canonical transformation of
the quadrature variables $\xi\rightarrow \xi^{'} = S\xi$,
Wigner distribution changes as $S: W(\xi)\rightarrow
W(S^{-1}\xi)$.  Consequently, under BS mixing input Wigner
distribution  $W_{A_{1}A_{2}F_{0}}(\xi_{1},\xi_{2},\xi_{3})$
changes as,

\begin{widetext} 
\begin{align}
W_{A_{1}A_{2}F_{0}}(\xi_{1},\xi_{2},\xi_{3}) &
\xrightarrow{\rm{BS}}
W_{A_{1}A^{'}_{2}F_{1}}(\xi_{1},\xi_{2},\xi_{3}) \nonumber
\\ &=8 e^{-d^{2}} e^{-\frac{\mu^{2} + \nu^{2}}{2}(x_{1}^{2}
+ p_{1}^{2}) + d(\mu - \nu)x_{1}} e^{-\frac{\mu^{2} -
(1-2\tau)\nu^{2}}{2}(x_{2}^{2} + p_{2}^{2}) +
2\mu\nu\sqrt{\tau} (x_{1}x_{2} - p_{1}p_{2}) +
d\sqrt{\tau}(\mu - \nu)x_{2}}~ \times \nonumber \\ &~~~~~~~~
e^{-\frac{\mu^{2} + (1-2\tau)\nu^{2}}{2}(x_{3}^{2} +
p_{3}^{2}) + \left( 2\nu\sqrt{1-\tau}( \nu\sqrt{\tau}x_{2} -
\mu x_{1}) -d\sqrt{1-\tau}(\mu - \nu)\right) x_{3} +
2\nu\sqrt{1-\tau}( \nu\sqrt{\tau}p_{2} + \mu p_{1}) p_{3}}.
\label{append_eq_exp_WigDist_mix} 
\end{align} 
\end{widetext}
The Wigner distribution for $k$-PSTMSC can then be written
as, 
\begin{widetext} 
\begin{align}
\tilde{W}_{A_{1}A^{'}_{2}}^{k}(\xi_{1},\xi_{2}) &
=\int\frac{dx_{3}dp_{3}}{4\pi}~
W_{A_{1}A^{'}_{2}F_{1}}(\xi_{1},\xi_{2},\xi_{3}) ~
W_{F_{1}}^{k}(\xi_{3}) \nonumber \\ 
&= 16 \frac{(-1)^{k}}{k!} e^{-d^{2}} e^{-\frac{\mu^{2} +
\nu^{2}}{2}(x_{1}^{2} + p_{1}^{2}) + d(\mu - \nu)x_{1}}
e^{-\frac{\mu^{2} - (1-2\tau)\nu^{2}}{2}(x_{2}^{2} +
p_{2}^{2}) + 2\mu\nu\sqrt{\tau} (x_{1}x_{2} - p_{1}p_{2}) +
d\sqrt{\tau}(\mu - \nu)x_{2}}~
\partial^{k}_{\eta}\partial^{k}_{\zeta} \Big[ e^{\eta\zeta}
~ \times \nonumber \\ &~~~ \int\frac{dx_{3}dp_{3}}{4\pi}
e^{-(\mu^{2} - \tau\nu^{2})(x_{3}^{2} + p_{3}^{2}) + \left(
(\eta - \zeta) + 2\nu\sqrt{1-\tau}( \nu\sqrt{\tau}x_{2} -
\mu x_{1}) -d\sqrt{1-\tau}(\mu - \nu)\right) x_{3} + \left(
i(\eta + \zeta) + 2\nu\sqrt{1-\tau}(\nu\sqrt{\tau}p_{2} +
\mu p_{1}) \right) p_{3} } \Big]_{\substack{\eta=0\\
\zeta=0}} \nonumber \\ &= \frac{4 (-1)^{k}}{k! (\mu^{2} -
T\nu^{2})} e^{-d^{2}} e^{-\frac{\mu^{2} +
\nu^{2}}{2}(x_{1}^{2} + p_{1}^{2}) + d(\mu - \nu)x_{1}}
e^{-\frac{\mu^{2} - (1-2\tau)\nu^{2}}{2}(x_{2}^{2} +
p_{2}^{2}) + 2\mu\nu\sqrt{\tau} (x_{1}x_{2} - p_{1}p_{2}) +
d\sqrt{\tau}(\mu - \nu)x_{2}}~ \times \nonumber \\ &~~~~~
\partial^{k}_{\eta}\partial^{k}_{\zeta} \left[ e^{\eta\zeta}
e^{\frac{1}{4(\mu^{2} - \tau\nu^{2})} \left\lbrace \left(
(\eta - \zeta) + 2\nu\sqrt{1-\tau}(\nu\sqrt{\tau}x_{2} - \mu
x_{1}) - d\sqrt{1-\tau}(\mu - \nu)\right)^{2} + \left(
i(\eta + \zeta) + 2\nu\sqrt{1-\tau}(\nu\sqrt{\tau}p_{2} +
\mu p_{1}) \right)^{2} \right\rbrace}
\right]_{\substack{\eta=0\\ \zeta=0}} \nonumber \\ &=
\frac{4 (-1)^{k}}{k! (\mu^{2} - \tau\nu^{2})} e^{-d^{2}}
e^{-\frac{\mu^{2} + \nu^{2}}{2}(x_{1}^{2} + p_{1}^{2}) +
d(\mu - \nu)x_{1}} e^{-\frac{\mu^{2} -
(1-2\tau)\nu^{2}}{2}(x_{2}^{2} + p_{2}^{2}) +
2\mu\nu\sqrt{\tau} (x_{1}x_{2} - p_{1}p_{2}) +
d\sqrt{\tau}(\mu - \nu)x_{2}}~ \times \nonumber \\ &~~~~~
e^{\frac{\nu^{2}(1-\tau)}{\mu^{2} - \tau\nu^{2}}
\left\lbrace (\nu\sqrt{\tau}x_{2} - \mu x_{1})^{2} +
(\nu\sqrt{\tau}p_{2} + \mu p_{1})^{2} \right\rbrace}
e^{\frac{d^{2}(1-\tau)(\mu - \nu)^{2}}{4(\mu^{2} -
\tau\nu^{2})}} e^{-\frac{d\nu(\mu - \nu)(1-\tau)}{\mu^{2} -
\tau\nu^{2}}(\nu\sqrt{\tau}x_{2} - \mu x_{1})}~ \times
\nonumber \\ &~~~~~ \partial^{k}_{\eta}\partial^{k}_{\zeta}
\left[ e^{\frac{\nu^{2}(1 - \tau)}{\mu^{2} - \tau\nu^{2}}
\eta\zeta +  \frac{\sqrt{1-\tau}}{\mu^{2} - \tau\nu^{2}}
\left\lbrace \nu^{2}\sqrt{\tau}(x_{2} + i p_{2}) - \mu\nu
(x_{1} - i p_{1}) - \frac{d(\mu - \nu)}{2} \right\rbrace
\eta -\frac{\sqrt{1-\tau}}{\mu^{2} - \tau\nu^{2}}
\left\lbrace \nu^{2}\sqrt{\tau}(x_{2} - i p_{2}) - \mu\nu
(x_{1} + i p_{1}) - \frac{d(\mu - \nu)}{2} \right\rbrace
\zeta } \right]_{\substack{\eta=0\\ \zeta=0}} \nonumber \\
&=(-A)^{k} W_{A_{1}A_{2}^{'}}^{0}(\xi_{1},\xi_{2})~  L_{k}
\left( \frac{|\xi_{12}|^{2}}{\nu^{2}(\mu^{2} - \tau\nu^{2})}
\right), \label{eq_exp_WigDist_k-PSTMSC} 
\end{align} 
\end{widetext}
where,
\begin{widetext}
\begin{align} 
A &=  \frac{\nu^{2}(1 - \tau)}{(\mu^{2} -
\tau\nu^{2})} ~;~ \xi_{12} = \nu^{2}\sqrt{\tau}(x_{2} + i
p_{2}) - \mu\nu (x_{1} - i p_{1}) - \frac{d(\mu - \nu)}{2}
~~~~~ \rm{and} \nonumber \\
W_{A_{1}A_{2}^{'}}^{0}(\xi_{1},\xi_{2}) &= \frac{4}{\mu^{2}
- \tau\nu^{2}} e^{-d^{2}} e^{-\frac{\mu^{2} +
\nu^{2}}{2}(x_{1}^{2} + p_{1}^{2}) -\frac{\mu^{2} -
(1-2\tau)\nu^{2}}{2}(x_{2}^{2} + p_{2}^{2}) +
2\mu\nu\sqrt{\tau} (x_{1}x_{2} - p_{1}p_{2}) + d(\mu -
\nu)(x_{1} + \sqrt{\tau}x_{2})} e^{\frac{1-\tau}{\mu^{2} -
\tau\nu^{2}} |\xi_{12}|^{2}}.
\label{append_eq_exp_WigDist_k-PSTMSC_unnorm} 
\end{align}
\end{widetext}
%%%%%%%%%%%%%%%
\section{Probability for $k$-PSTMSC}
\label{sec_append:Prob. k-PSTMSC}

In this section we calculate the probability of successfully
subtracting $k$ photons from a TMSC state. This probability
will also serve as a normalization to the Wigner
distribution as derived above.

Probability of $k$-photon subtraction is obtained by
integrating the resultant Wigner distribution, i.e., $P(k)
=\int \frac{dx_{1}dp_{1}}{4\pi}
\int\frac{dx_{2}dp_{2}}{4\pi}
W_{A_{1}A_{2}^{'}}^{k}(\xi_{1},\xi_{2})$.  In a straightforward
calculation it could be shown that 
\begin{widetext}
\begin{align} 
P^{k}_{PS} &= (-A)^{k} \int
\frac{dx_{1}dp_{1}}{4\pi} \int\frac{dx_{2}dp_{2}}{4\pi}
W_{A_{1}A_{2}^{'}}^{0}(\xi_{1},\xi_{2})~  L_{k} \left(
\frac{|\xi_{12}|^{2}}{\nu^{2}(\mu^{2} - \tau\nu^{2})}
\right) \nonumber \\ &=   \frac{(-1)^{k}}{k! (\mu^{2} -
\tau\nu^{2})} e^{- \left( 1 - \frac{(1-\tau)(\mu -
\nu)^{2}}{4(\mu^{2} - \tau\nu^{2})}\right)d^{2}}
\partial^{k}_{\eta}\partial^{k}_{\zeta} \Big[
e^{\frac{\nu^{2}(1 - \tau)}{\mu^{2} - \tau\nu^{2}} \eta\zeta
- \frac{d(\mu - \nu)\sqrt{1 - \tau}}{2(\mu^{2} -
\tau\nu^{2})} (\eta - \zeta)}  ~\times \nonumber \\ &~~~
\int\frac{dx_{1}dp_{1}}{2\pi} e^{-\frac{\mu^{2} +
\tau\nu^{2}}{2(\mu^{2} - \tau\nu^{2})}(x_{1}^{2}+p_{1}^{2})
+ \frac{1}{\mu^{2} - \tau\nu^{2}} \left\lbrace d(\mu -
\tau\nu) - \mu\nu\sqrt{1 - \tau}(\eta - \zeta) \right\rbrace
x_{1} + \frac{i\mu\nu\sqrt{1 - \tau}}{\mu^{2} - \tau\nu^{2}}
(\eta + \zeta)p_{1} } ~\times \nonumber \\ &~~~
\int\frac{dx_{2}}{\sqrt{2\pi}}  e^{-\frac{\mu^{2} +
\tau\nu^{2}}{2(\mu^{2} - \tau\nu^{2})}x_{2}^{2} +
\frac{\sqrt{\tau}}{\mu^{2} - \tau\nu^{2}} \left\lbrace
2\mu\nu x_{1} + \nu^{2}\sqrt{1 - \tau}(\eta - \zeta) + d(\mu
- \nu) \right\rbrace x_{2}} \int\frac{dp_{2}}{\sqrt{2\pi}}
  e^{-\frac{\mu^{2} + \tau\nu^{2}}{2(\mu^{2} -
\tau\nu^{2})}p_{2}^{2} - \frac{\sqrt{\tau}}{\mu^{2} -
\tau\nu^{2}} \left\lbrace 2\mu\nu p_{1} -  i\nu^{2}\sqrt{1 -
\tau}(\eta + \zeta) \right\rbrace p_{2}}
\Big]_{\substack{\eta=0\\ \zeta=0}} 
\nonumber 
\\
&=  \frac{(-1)^{k}}{k! (\mu^{2} - T\nu^{2})}  \frac{\mu^{2}
- \tau\nu^{2}}{\mu^{2} + \tau\nu^{2}}
e^{- \left\lbrace 1 - \frac{(1-\tau)(\mu -
\nu)^{2}}{4(\mu^{2} - \tau\nu^{2})} - \frac{(\mu - \nu)^{2}
\tau}{2(\mu^{4} - \tau^{2}\nu^{4})} \right\rbrace d^{2}}
\partial^{k}_{\eta}\partial^{k}_{\zeta} \Big[  
e^{\frac{\nu^{2}(1 - \tau)}{\mu^{2} + \tau\nu^{2}} \eta\zeta
- \frac{d(\mu - \nu)\sqrt{1 - \tau}}{2(\mu^{2} +
\tau\nu^{2})} (\eta - \zeta)}  ~\times
\nonumber
\\
&~~~ \int\frac{dx_{1}}{\sqrt{2\pi}} e^{-\frac{\mu^{2} -
\tau\nu^{2}}{2(\mu^{2} + \tau\nu^{2})}x_{1}^{2} +
\frac{1}{\mu^{2} + \tau\nu^{2}} \left\lbrace d(\mu +
\tau\nu) - \mu\nu\sqrt{1 - \tau}(\eta - \zeta) \right\rbrace
x_{1} } 
\int\frac{dp_{1}}{\sqrt{2\pi}} e^{-\frac{\mu^{2} -
\tau\nu^{2}}{2(\mu^{2} + \tau\nu^{2})} p_{1}^{2} +
\frac{i\mu\nu\sqrt{1 - \tau}}{\mu^{2} + \tau\nu^{2}} (\eta +
\zeta)p_{1} }
 \Big]_{\substack{\eta=0\\ \zeta=0}}
\nonumber
\\
&= \frac{(-1)^{k}}{k! (\mu^{2} - \tau\nu^{2})} e^{-
\left\lbrace 1 - \frac{(1-\tau)(\mu - \nu)^{2}}{4(\mu^{2} -
\tau\nu^{2})} - \frac{(\mu - \nu)^{2} \tau}{2(\mu^{4} -
\tau^{2}\nu^{4})} - \frac{(\mu + \tau\nu)^{2}}{2(\mu^{4} -
\tau^{2}\nu^{4})} \right\rbrace d^{2}}
~\times
\partial^{k}_{\eta}\partial^{k}_{\zeta} \left[  
e^{-\frac{\nu^{2}(1 - \tau)}{\mu^{2} - \tau\nu^{2}} \eta\zeta 
- \frac{d(\mu + \nu)\sqrt{1 - \tau}}{2(\mu^{2} - \tau\nu^{2})} \eta 
+ \frac{d(\mu + \nu)\sqrt{1 - \tau}}{2(\mu^{2} - \tau\nu^{2})} \zeta } 
\right]_{\substack{\eta=0\\ \zeta=0}} 
\nonumber
\\
&= \frac{A^{k}}{\mu^{2} - \tau\nu^{2}}  e^{-  \frac{(1-\tau)(\mu + \nu)^{2}}{4(\mu^{2} - \tau\nu^{2})} d^{2}}  L_{k}\left( -\frac{d^{2}(\mu + \nu)^{2}}{4 \nu^{2} (\mu^{2} - \tau\nu^{2})} \right).
\label{append_eq_exp_prob_k-PSTMSC}
\end{align}
\end{widetext}

Thus the normalized Wigner distribution for $k$-PSTMSC is given by
\begin{equation}
W_{A_{1}A^{'}_{2}}^{k}(\xi_{1},\xi_{2}) = \left( P^{k}_{PS} \right)^{-1} \tilde{W}_{A_{1}A^{'}_{2}}^{k}(\xi_{1},\xi_{2})
\label{append_eq_exp_WigDist_k-PSTMSC_norm}
\end{equation} 

\section{Co-variance Matrix for $k$-PSTMSC}
\label{sec_append:Covariance Matrix k-PSTMSC}

In this section we derive the co-variance matrix for the $k$-PSTMSC state by using moment generating functions.

The moment generating function for the $k$-PSTMSC is given by 
\begin{widetext}
\begin{align}
C_{i,j}^{m,n} &= \left\langle x_{1}^{i} p_{1}^{j} x_{2}^{m} p_{2}^{n} \right\rangle = \frac{1}{P^{k}_{PS}} \int\frac{dx_{1}dp_{1}}{4\pi} \int\frac{dx_{2}dp_{2}}{4\pi} \tilde{W}_{A_{1}A_{2}^{'}}^{k}(\xi_{1},\xi_{2}) x_{1}^{i} p_{1}^{j} x_{2}^{m} p_{2}^{n}
\nonumber
\\
&= \frac{1}{P^{k}_{PS}} \partial_{a}^{i} \partial_{b}^{j} \partial_{s}^{m} \partial_{t}^{n} \left[ 
\int\frac{dx_{1}dp_{1}}{4\pi} \int\frac{dx_{2}dp_{2}}{4\pi} \tilde{W}_{A_{1}A_{2}^{'}}^{k}(\xi_{1},\xi_{2})
e^{a x_{1} + b p_{1} + s x_{2} + t p_{2}}
\right]_{\substack{a=0,b=0\\ s=0,t=0}}
\nonumber
\\
&= \frac{1}{P^{k}_{PS}} \frac{(-1)^{k}}{k! (\mu^{2} - \tau\nu^{2})} e^{- \left( 1 - \frac{(1-\tau)(\mu - \nu)^{2}}{4(\mu^{2} - \tau\nu^{2})}\right)d^{2}}
\partial_{a}^{i} \partial_{b}^{j} \partial_{s}^{m} \partial_{t}^{n} \partial^{k}_{\eta} \partial^{k}_{\zeta} \Big[  e^{\frac{\nu^{2}(1 - \tau)}{\mu^{2} - \tau\nu^{2}} \eta\zeta - \frac{d(\mu - \nu)\sqrt{1 - \tau}}{2(\mu^{2} - \tau\nu^{2})} (\eta - \zeta)}  ~\times
\nonumber
\\
&~~~  \int\frac{dx_{1}dp_{1}}{2\pi} e^{-\frac{\mu^{2} + \tau\nu^{2}}{2(\mu^{2} - \tau\nu^{2})}(x_{1}^{2}+p_{1}^{2}) + \frac{1}{\mu^{2} - \tau\nu^{2}} \left\lbrace d(\mu - \tau\nu) - \mu\nu\sqrt{1 - \tau}(\eta - \zeta) + a (\mu^{2} - \tau\nu^{2}) \right\rbrace x_{1} + \left\lbrace \frac{i\mu\nu\sqrt{1 - \tau}}{\mu^{2} - \tau\nu^{2}} (\eta + \zeta) +b \right\rbrace p_{1} } ~\times
\nonumber
\\
&~~~~ \int\frac{dx_{2}}{\sqrt{2\pi}}
e^{-\frac{\mu^{2} + \tau\nu^{2}}{2(\mu^{2} - \tau\nu^{2})}x_{2}^{2} + \frac{\sqrt{\tau}}{\mu^{2} - \tau\nu^{2}} \left\lbrace 2\mu\nu x_{1} + \nu^{2}\sqrt{1 - \tau}(\eta - \zeta) + d(\mu - \nu) + s\frac{\mu^{2} - \tau\nu^{2}}{\sqrt{\tau}} \right\rbrace x_{2}}  
\nonumber
\\ 
&~~~ \int\frac{dp_{2}}{\sqrt{2\pi}}
e^{-\frac{\mu^{2} + \tau\nu^{2}}{2(\mu^{2} - \tau\nu^{2})}p_{2}^{2} - 
\frac{\sqrt{\tau}}{\mu^{2} - \tau\nu^{2}} \left\lbrace 2\mu\nu p_{1} -  i\nu^{2}\sqrt{1 - \tau}(\eta + \zeta) - t\frac{\mu^{2} - \tau\nu^{2}}{\sqrt{\tau}} \right\rbrace p_{2}}  
\Big]_{\substack{a=0,b=0,\eta=0\\ s=0,t=0,\zeta=0}}
\nonumber
\\
&= \frac{1}{P^{k}_{PS}}  \frac{(-1)^{k}}{k! (\mu^{2} + \tau\nu^{2})}  
e^{- \left\lbrace 1 - \frac{(1-\tau)(\mu - \nu)^{2}}{4(\mu^{2} - \tau\nu^{2})} - \frac{(\mu - \nu)^{2} \tau}{2(\mu^{4} - \tau^{2}\nu^{4})} \right\rbrace d^{2}}  
\partial_{a}^{i} \partial_{b}^{j} \partial_{s}^{m} \partial_{t}^{n} \Big[ 
e^{\frac{d(\mu - \nu)\sqrt{\tau}}{\mu^{2} + \tau\nu^{2}}c + \frac{\mu^{2} - \tau\nu^{2}}{2(\mu^{2} + \tau\nu^{2})} \left(s^{2} + t^{2}\right)} ~\times
\nonumber
\\
&~~~  \partial^{k}_{\eta} \partial^{k}_{\zeta} \Big[  
e^{\frac{\nu^{2}(1 - \tau)}{\mu^{2} + \tau\nu^{2}} \eta\zeta 
 - \left\lbrace \frac{d(\mu - \nu)\sqrt{1 - \tau}}{2(\mu^{2} + \tau\nu^{2})} - \frac{\nu^{2}\sqrt{\tau(1 - \tau)}}{\mu^{2} + \tau\nu^{2}}s \right\rbrace (\eta - \zeta)
+ \frac{i\nu^{2}t\sqrt{\tau(1 - \tau)}}{\mu^{2} + \tau\nu^{2}} (\eta + \zeta)} ~\times
\nonumber
\\
&~~~  \int\frac{dx_{1}}{\sqrt{2\pi}} e^{-\frac{\mu^{2} - \tau\nu^{2}}{2(\mu^{2} + \tau\nu^{2})}x_{1}^{2} + 
\frac{1}{\mu^{2} + \tau\nu^{2}} \left\lbrace d(\mu + \tau\nu) - \mu\nu\sqrt{1 - \tau}(\eta - \zeta) + a(\mu^{2} + \tau\nu^{2}) + 2\mu\nu\sqrt{\tau}s \right\rbrace x_{1} } 
\times
\nonumber
\\
&~~~~  \int\frac{dp_{1}}{\sqrt{2\pi}} e^{-\frac{\mu^{2} - \tau\nu^{2}}{2(\mu^{2} + \tau\nu^{2})} p_{1}^{2} + 
\frac{\mu\nu}{\mu^{2} + \tau\nu^{2}} \left\lbrace i\sqrt{1 - \tau}(\eta + \zeta) - 2\sqrt{\tau}t + b\frac{\mu^{2} + \tau\nu^{2}}{\mu\nu} \right\rbrace p_{1} }
\Big]_{\substack{\eta=0\\ \zeta=0}}
\Big]_{\substack{a=0,b=0\\ s=0,t=0}}
\nonumber
\\
&= \frac{1}{P^{k}_{PS}}  \frac{(-1)^{k}}{k! (\mu^{2} - \tau\nu^{2})} 
e^{-  \frac{(1-\tau)(\mu + \nu)^{2}}{4(\mu^{2} - \tau\nu^{2})} d^{2}}   ~\times
\partial_{a}^{i} \partial_{b}^{j} \partial_{s}^{m} \partial_{t}^{n} \Big[
e^{ \frac{\mu^{2} + \tau\nu^{2}}{2(\mu^{2} - \tau\nu^{2})} \left( a^{2} + b^{2} + s^{2} + t^{2} \right) + \frac{2\mu\nu\sqrt{\tau}}{\mu^{2} - \tau\nu^{2}} (a s - b t)}
~\times
\nonumber
\\
&~~~   e^{ \frac{\mu + \tau\nu}{\mu^{4} - \tau^{2}\nu^{4}}\left\lbrace a(\mu^{2} + \tau\nu^{2}) + 2\mu\nu\sqrt{\tau}s \right\rbrace d}  ~\times
\partial^{k}_{\eta} \partial^{k}_{\zeta} \left[
e^{-\frac{\nu^{2}(1 - \tau)}{\mu^{2} - \tau\nu^{2}}\eta\zeta} 
e^{- \frac{Z\sqrt{1 - \tau}}{2(\mu^{2} - \tau\nu^{2})} \eta 
+ \frac{Z^{*}\sqrt{1 - \tau}}{2(\mu^{2} - \tau\nu^{2})} \zeta }
\right]_{\substack{\eta=0\\ \zeta=0}} 
\Big]_{\substack{a=0,b=0\\ s=0,t=0}}
\nonumber
\\
&= \frac{1}{L_{k}\left( -\frac{d^{2}(\mu + \nu)^{2}}{4 \nu^{2} (\mu^{2} - \tau\nu^{2})} \right)}
 \partial_{a}^{i} \partial_{b}^{j} \partial_{s}^{m} \partial_{t}^{n} \Big[ 
 e^{ \frac{\mu^{2} + \tau\nu^{2}}{2(\mu^{2} - \tau\nu^{2})} \left( a^{2} + b^{2} + s^{2} + t^{2} \right) + \frac{2\mu\nu\sqrt{\tau}}{\mu^{2} - \tau\nu^{2}} (a s - b t) 
+  \frac{\mu + \tau\nu}{\mu^{4} - \tau^{2}\nu^{4}}\left\lbrace a(\mu^{2} + \tau\nu^{2}) + 2\mu\nu\sqrt{\tau}s \right\rbrace d} ~\times
\nonumber
\\
&~~~~~~~~~~~~~~~~~~~~~~~~~~~~~~~~~~~~~~~~~~~~  L_{k} \left( -\frac{|Z|^{2}}{4\nu^{2}(\mu^{2} - \tau\nu^{2})} \right)
\Big]_{\substack{a=0,b=0\\ s=0,t=0}},
\label{eq_exp_mgf_k-PSTMSC}
\end{align}
\end{widetext}
where
$Z=d(\mu + \nu) + 2\mu\nu (a - i b) + 2\sqrt{\tau}\nu^{2} (s + i t)$. 
We express various moments, e.g., $\langle x_{1}^{l} \rangle$ in a compact notation as $\langle x_{1}^{i} \rangle = C_{i,j}^{m,n} \delta_{i,l}\delta_{j,0}\delta_{m,0}\delta_{n,0}$.
This easily leads to,
\begin{widetext}
\begin{subequations}
\begin{align}
\left\langle x_{1} \right\rangle &= C_{i,j}^{m,n} \delta_{i,1}\delta_{j,0}\delta_{m,0}\delta_{n,0}
\nonumber
\\
&= \frac{1}{L_{k}\left( -\frac{d^{2}(\mu + \nu)^{2}}{4 \nu^{2} (\mu^{2} - \tau\nu^{2})} \right)}
 \partial_{a} \left[ 
 e^{ \frac{\mu^{2} + \tau\nu^{2}}{2(\mu^{2} - \tau\nu^{2})} a^{2} 
+  \frac{d(\mu + \tau\nu)}{\mu^{2} - \tau\nu^{2}} a} ~\times
 L_{k} \left( -\frac{ ( d(\mu + \nu) + 2\mu\nu a )^{2}}{4\nu^{2}(\mu^{2} - \tau\nu^{2})} \right)
\right]_{a=0}
\nonumber
\\
&= \frac{d(\mu + \tau\nu)}{\mu^{2} - \tau\nu^{2}} 
+ \frac{d\mu(\mu + \nu)}{\nu(\mu^{2} - \tau\nu^{2})} \frac{ L_{k-1}^{1} \left( -\frac{ d^{2}(\mu + \nu)^{2}}{4\nu^{2}(\mu^{2} - \tau\nu^{2})} \right)}{L_{k}\left( -\frac{d^{2}(\mu + \nu)^{2}}{4 \nu^{2} (\mu^{2} - \tau\nu^{2})} \right)}
\end{align}

\begin{align}
\left\langle x_{1}^{2} \right\rangle &= C_{i,j}^{m,n} \delta_{i,2}\delta_{j,0}\delta_{m,0}\delta_{n,0}
\nonumber
\\
&= \frac{1}{L_{k}\left( -\frac{d^{2}(\mu + \nu)^{2}}{4 \nu^{2} (\mu^{2} - \tau\nu^{2})} \right)}
 \partial_{a}^{2} \left[ 
 e^{ \frac{\mu^{2} + \tau\nu^{2}}{2(\mu^{2} - \tau\nu^{2})} a^{2} 
+  \frac{d(\mu + \tau\nu)}{\mu^{2} - \tau\nu^{2}} a} ~\times
 L_{k} \left( -\frac{ ( d(\mu + \nu) + 2\mu\nu a )^{2}}{4\nu^{2}(\mu^{2} - \tau\nu^{2})} \right)
\right]_{a=0}
\nonumber
\\
&=\left( \frac{d(\mu + \tau\nu)}{\mu^{2} - \tau\nu^{2}} \right)^{2}
+ \frac{\mu^{2} + \tau\nu^{2}}{\mu^{2} - \tau\nu^{2}}
+ 2 \left\lbrace \frac{d^{2}\mu(\mu + \nu)(\mu + \tau\nu)}{\nu(\mu^{2} - \tau\nu^{2})^{2}} + \frac{\mu^{2}}{\mu^{2} - \tau\nu^{2}} \right\rbrace  \frac{ L_{k-1}^{1} \left( -\frac{ d^{2}(\mu + \nu)^{2}}{4\nu^{2}(\mu^{2} - \tau\nu^{2})} \right)}{L_{k}\left( -\frac{d^{2}(\mu + \nu)^{2}}{4 \nu^{2} (\mu^{2} - \tau\nu^{2})} \right)} ~+
\nonumber
\\
&~~~~~  \frac{d^{2}\mu^{2}(\mu + \nu)^{2}}{\nu^{2}(\mu^{2} - \tau\nu^{2})^{2}} \frac{ L_{k-2}^{2} \left( -\frac{ d^{2}(\mu + \nu)^{2}}{4\nu^{2}(\mu^{2} - \tau\nu^{2})} \right)}{L_{k}\left( -\frac{d^{2}(\mu + \nu)^{2}}{4 \nu^{2} (\mu^{2} - \tau\nu^{2})} \right)}
\end{align}

\begin{align}
V_{A}^{x} &= \left\langle x_{1}^{2} \right\rangle - \left\langle x_{1} \right\rangle^{2}
\nonumber
\\
&= \frac{\mu^{2} + \tau\nu^{2}}{\mu^{2} - \tau\nu^{2}}
+ \frac{2\mu^{2}}{\mu^{2} - \tau\nu^{2}} \frac{ L_{k-1}^{1} \left( -\frac{ d^{2}(\mu + \nu)^{2}}{4\nu^{2}(\mu^{2} - \tau\nu^{2})} \right)}{L_{k}\left( -\frac{d^{2}(\mu + \nu)^{2}}{4 \nu^{2} (\mu^{2} - \tau\nu^{2})} \right)} ~+
\frac{d^{2}\mu^{2}(\mu + \nu)^{2}}{\nu^{2}(\mu^{2} - \tau\nu^{2})^{2}} \Bigg\lbrace  
\frac{ L_{k-2}^{2} \left( -\frac{ d^{2}(\mu + \nu)^{2}}{4\nu^{2}(\mu^{2} - \tau\nu^{2})} \right)}{L_{k}\left( -\frac{d^{2}(\mu + \nu)^{2}}{4 \nu^{2} (\mu^{2} - \tau\nu^{2})} \right)}
\nonumber
\\
&~~~~~~~~~~~~~~~~~~~~~~~~~~~~~~~~~~~~~~~~~~~~~~~~~~~~~~~~~~~~~~~~ 
~~~~~~~~~~~~ - \left( \frac{ L_{k-1}^{1} \left( -\frac{ d^{2}(\mu + \nu)^{2}}{4\nu^{2}(\mu^{2} - \tau\nu^{2})} \right)}{L_{k}\left( -\frac{d^{2}(\mu + \nu)^{2}}{4 \nu^{2} (\mu^{2} - \tau\nu^{2})} \right)} \right)^{2}
\Bigg\rbrace
\end{align}
\end{subequations}

\begin{subequations}
\begin{align}
\left\langle p_{1} \right\rangle &= C_{i,j}^{m,n} \delta_{i,0}\delta_{j,1}\delta_{m,0}\delta_{n,0}
\nonumber
\\
&= \frac{1}{L_{k}\left( -\frac{d^{2}(\mu + \nu)^{2}}{4 \nu^{2} (\mu^{2} - \tau\nu^{2})} \right)}
 \partial_{b} \left[ 
 e^{ \frac{\mu^{2} + \tau\nu^{2}}{2(\mu^{2} - \tau\nu^{2})} b^{2}} ~\times
 L_{k} \left( -\frac{ | d(\mu + \nu) - 2i\mu\nu b |^{2}}{4\nu^{2}(\mu^{2} - \tau\nu^{2})} \right)
\right]_{b=0}=0
\end{align}

\begin{align}
V_{A}^{p} = \left\langle p_{1}^{2} \right\rangle &= C_{i,j}^{m,n} \delta_{i,0}\delta_{j,2}\delta_{m,0}\delta_{n,0} 
\nonumber
\\
&= \frac{1}{L_{k}\left( -\frac{d^{2}(\mu + \nu)^{2}}{4 \nu^{2} (\mu^{2} - \tau\nu^{2})} \right)}
 \partial_{b}^{2} \left[ 
 e^{ \frac{\mu^{2} + \tau\nu^{2}}{2(\mu^{2} - \tau\nu^{2})} a^{2}} ~\times
 L_{k} \left( -\frac{ | d(\mu + \nu) - 2i\mu\nu b |^{2}}{4\nu^{2}(\mu^{2} - \tau\nu^{2})} \right)
\right]_{b=0}
\nonumber
\\
&= \frac{\mu^{2} + \tau\nu^{2}}{\mu^{2} - \tau\nu^{2}} 
+\frac{2\mu^{2}}{\mu^{2} - \tau\nu^{2}} \frac{ L_{k-1}^{1} \left( -\frac{ d^{2}(\mu + \nu)^{2}}{4\nu^{2}(\mu^{2} - \tau\nu^{2})} \right)}{L_{k}\left( -\frac{d^{2}(\mu + \nu)^{2}}{4 \nu^{2} (\mu^{2} - \tau\nu^{2})} \right)}
\end{align}
\end{subequations}

\begin{subequations}
\begin{align}
\left\langle x_{2} \right\rangle &= C_{i,j}^{m,n} \delta_{i,0}\delta_{j,0}\delta_{m,1}\delta_{n,0}
\nonumber
\\
&= \frac{1}{L_{k}\left( -\frac{d^{2}(\mu + \nu)^{2}}{4 \nu^{2} (\mu^{2} - \tau\nu^{2})} \right)}
 \partial_{s} \left[ 
 e^{ \frac{\mu^{2} + \tau\nu^{2}}{2(\mu^{2} - \tau\nu^{2})} s^{2} 
+  \frac{2 d \mu\nu(\mu + \tau\nu)\sqrt{T}}{\mu^{4} - \tau^{2}\nu^{4}} s} ~\times
 L_{k} \left( -\frac{ ( d(\mu + \nu) + 2\nu^{2}\sqrt{\tau} s )^{2}}{4\nu^{2}(\mu^{2} - \tau\nu^{2})} \right)
\right]_{c=0}
\nonumber
\\
&= \frac{2d\mu\nu(\mu + \tau\nu)\sqrt{\tau}}{\mu^{4} - \tau^{2}\nu^{4}}
+ \frac{d(\mu + \nu)}{(\mu^{2} - \tau\nu^{2})} \frac{ L_{k-1}^{1} \left( -\frac{ d^{2}(\mu + \nu)^{2}}{4\nu^{2}(\mu^{2} - \tau\nu^{2})} \right)}{L_{k}\left( -\frac{d^{2}(\mu + \nu)^{2}}{4 \nu^{2} (\mu^{2} - \tau\nu^{2})} \right)}
\end{align}

\begin{align}
\left\langle x_{2}^{2} \right\rangle &= C_{i,j}^{m,n} \delta_{i,0}\delta_{j,0}\delta_{m,2}\delta_{n,0}
\nonumber
\\
&= \frac{1}{L_{k}\left( -\frac{d^{2}(\mu + \nu)^{2}}{4 \nu^{2} (\mu^{2} - \tau\nu^{2})} \right)}
 \partial_{s}^{2} \left[ 
 e^{ \frac{\mu^{2} + \tau\nu^{2}}{2(\mu^{2} - \tau\nu^{2})} s^{2} 
+  \frac{2d\mu\nu(\mu + \tau\nu)\sqrt{\tau}}{\mu^{4} - \tau^{2}\nu^{4}} s} ~\times
 L_{k} \left( -\frac{ ( d(\mu + \nu) + 2\nu^{2}\sqrt{\tau} s )^{2}}{4\nu^{2}(\mu^{2} - \tau\nu^{2})} \right)
\right]_{c=0}
\nonumber
\\
&=\left( \frac{2d\mu\nu(\mu + \tau\nu)\sqrt{\tau}}{\mu^{4} - \tau^{2}\nu^{4}}  \right)^{2}
+ \frac{\mu^{2} + \tau\nu^{2}}{\mu^{2} - \tau\nu^{2}}
+ 2 \left\lbrace \frac{2 d^{2}\mu\nu \tau(\mu + \nu)(\mu + \tau\nu)}{(\mu^{2} - \tau\nu^{2})(\mu^{4} - \tau^{2}\nu^{4})} + \frac{\nu^{2} \tau}{\mu^{2} - \tau\nu^{2}} \right\rbrace  \frac{ L_{k-1}^{1} \left( -\frac{ d^{2}(\mu + \nu)^{2}}{4\nu^{2}(\mu^{2} - \tau\nu^{2})} \right)}{L_{k}\left( -\frac{d^{2}(\mu + \nu)^{2}}{4 \nu^{2} (\mu^{2} - \tau\nu^{2})} \right)} ~+
\nonumber
\\
&~~~~~ \frac{d^{2}(\mu + \nu)^{2} \tau}{(\mu^{2} - \tau\nu^{2})^{2}} \frac{ L_{k-2}^{2} \left( -\frac{ d^{2}(\mu + \nu)^{2}}{4\nu^{2}(\mu^{2} - \tau\nu^{2})} \right)}{L_{k}\left( -\frac{d^{2}(\mu + \nu)^{2}}{4 \nu^{2} (\mu^{2} - \tau\nu^{2})} \right)}
\end{align}

\begin{align}
V_{B}^{x} &= \left\langle x_{2}^{2} \right\rangle - \left\langle x_{2} \right\rangle^{2}
\nonumber
\\
&= \frac{\mu^{2} + \tau\nu^{2}}{\mu^{2} - \tau\nu^{2}}
+ \frac{2\nu^{2} \tau}{\mu^{2} - \tau\nu^{2}} \frac{ L_{k-1}^{1} \left( -\frac{ d^{2}(\mu + \nu)^{2}}{4\nu^{2}(\mu^{2} - \tau\nu^{2})} \right)}{L_{k}\left( -\frac{d^{2}(\mu + \nu)^{2}}{4 \nu^{2} (\mu^{2} - \tau\nu^{2})} \right)} ~+
\frac{d^{2}(\mu + \nu)^{2} \tau}{(\mu^{2} - \tau\nu^{2})^{2}} \Bigg\lbrace  
\frac{ L_{k-2}^{2} \left( -\frac{ d^{2}(\mu + \nu)^{2}}{4\nu^{2}(\mu^{2} - \tau\nu^{2})} \right)}{L_{k}\left( -\frac{d^{2}(\mu + \nu)^{2}}{4 \nu^{2} (\mu^{2} - \tau\nu^{2})} \right)}
\nonumber
\\
&~~~~~~~~~~~~~~~~~~~~~~~~~~~~~~~~~~~~~~~~~~~~~~~~~~~~~~~~~~~~~~~~ 
~~~~~~~~~~ - \left( \frac{ L_{k-1}^{1} \left( -\frac{ d^{2}(\mu + \nu)^{2}}{4\nu^{2}(\mu^{2} - \tau\nu^{2})} \right)}{L_{k}\left( -\frac{d^{2}(\mu + \nu)^{2}}{4 \nu^{2} (\mu^{2} - \tau\nu^{2})} \right)} \right)^{2}
\Bigg\rbrace
\end{align}
\end{subequations}

\begin{subequations}
\begin{align}
\left\langle p_{2} \right\rangle &= C_{i,j}^{m,n} \delta_{i,0}\delta_{j,0}\delta_{m,0}\delta_{n,1}
\nonumber
\\
&= \frac{1}{L_{k}\left( -\frac{d^{2}(\mu + \nu)^{2}}{4 \nu^{2} (\mu^{2} - \tau\nu^{2})} \right)}
 \partial_{t} \left[ 
 e^{ \frac{\mu^{2} + \tau\nu^{2}}{2(\mu^{2} - \tau\nu^{2})} t^{2}} ~\times
 L_{k} \left( -\frac{ | d(\mu + \nu) + 2i\nu^{2}\sqrt{\tau} t |^{2}}{4\nu^{2}(\mu^{2} - \tau\nu^{2})} \right)
\right]_{d=0} = 0
\end{align}

\begin{align}
V_{B}^{p} = \left\langle p_{2}^{2} \right\rangle &= C_{i,j}^{m,n} \delta_{i,0}\delta_{j,0}\delta_{m,0}\delta_{n,2}
\nonumber
\\
&= \frac{1}{L_{k}\left( -\frac{d^{2}(\mu + \nu)^{2}}{4 \nu^{2} (\mu^{2} - \tau\nu^{2})} \right)}
 \partial_{t}^{2} \left[ 
 e^{ \frac{\mu^{2} + \tau\nu^{2}}{2(\mu^{2} - \tau\nu^{2})} t^{2}} ~\times
 L_{k} \left( -\frac{ | d(\mu + \nu) + 2i\nu^{2}\sqrt{\tau} t |^{2}}{4\nu^{2}(\mu^{2} - \tau\nu^{2})} \right)
\right]_{d=0}
\nonumber
\\
&= \frac{\mu^{2} + \tau\nu^{2}}{\mu^{2} - \tau\nu^{2}} 
+ \frac{2\nu^{2} \tau}{\mu^{2} - \tau\nu^{2}} \frac{ L_{k-1}^{1} \left( -\frac{ d^{2}(\mu + \nu)^{2}}{4\nu^{2}(\mu^{2} - \tau\nu^{2})} \right)}{L_{k}\left( -\frac{d^{2}(\mu + \nu)^{2}}{4 \nu^{2} (\mu^{2} - \tau\nu^{2})} \right)}
\end{align}
\end{subequations}

\begin{subequations}
\begin{align}
\left\langle x_{1}x_{2} \right\rangle &= C_{i,j}^{m,n} \delta_{i,1}\delta_{j,0}\delta_{m,1}\delta_{n,0}
\nonumber
\\
&= \frac{1}{L_{k}\left( -\frac{d^{2}(\mu + \nu)^{2}}{4 \nu^{2} (\mu^{2} - \tau\nu^{2})} \right)}
 \partial_{s} \Bigg[ 
 e^{ \frac{\mu^{2} + \tau\nu^{2}}{2(\mu^{2} - \tau\nu^{2})} s^{2} 
+  \frac{2d\mu\nu(\mu + \tau\nu)\sqrt{T}}{\mu^{4} - \tau^{2}\nu^{4}} s}
\partial_{a} \Bigg[ 
e^{ \frac{\mu^{2} + \tau\nu^{2}}{2(\mu^{2} - \tau\nu^{2})} a^{2} 
+  \frac{d(\mu + \tau\nu) + 2\mu\nu\sqrt{\tau} s}{\mu^{2} - \tau\nu^{2}} a} ~\times
\nonumber
\\
&~~~  L_{k} \left( -\frac{ ( d(\mu + \nu) + 2\mu\nu a + 2\nu^{2}\sqrt{\tau} s )^{2}}{4\nu^{2}(\mu^{2} - \tau\nu^{2})} \right)
\Bigg]_{a=0} 
\Bigg]_{c=0}
\nonumber
\\
&= \frac{1}{L_{k}\left( -\frac{d^{2}(\mu + \nu)^{2}}{4 \nu^{2} (\mu^{2} - \tau\nu^{2})} \right)}
\partial_{s} \Bigg[ 
 e^{ \frac{\mu^{2} + \tau\nu^{2}}{2(\mu^{2} - \tau\nu^{2})} s^{2} 
+  \frac{2d\mu\nu(\mu + \tau\nu)\sqrt{\tau}}{\mu^{4} - \tau^{2}\nu^{4}} s} 
\Bigg\lbrace  \frac{d(\mu + \tau\nu) + 2\mu\nu\sqrt{\tau} s}{\mu^{2} - \tau\nu^{2}} L_{k} \left( -\frac{ ( d(\mu + \nu) + 2\nu^{2}\sqrt{\tau} s )^{2}}{4\nu^{2}(\mu^{2} - \tau\nu^{2})} \right) + 
\nonumber
\\
&~~~  \frac{d\mu(\mu + \nu) + 2\mu\nu^{2}\sqrt{\tau} s}{\nu(\mu^{2} - \tau\nu^{2})} L_{k-1}^{1} \left( -\frac{ ( d(\mu + \nu) + 2\nu^{2}\sqrt{\tau} s )^{2}}{4\nu^{2}(\mu^{2} - \tau\nu^{2})} \right) 
\Bigg\rbrace
\Bigg]_{c=0}
\nonumber
\\
&= \frac{1}{L_{k}\left( -\frac{d^{2}(\mu + \nu)^{2}}{4 \nu^{2} (\mu^{2} - \tau\nu^{2})} \right)}
\Bigg[ 
\frac{2d\mu\nu(\mu + \tau\nu)\sqrt{\tau}}{\mu^{4} - \tau^{2}\nu^{4}}
\Bigg\lbrace  \frac{d(\mu + \tau\nu)}{\mu^{2} - \tau\nu^{2}} L_{k} \left( -\frac{ d^{2}(\mu + \nu)^{2}}{4\nu^{2}(\mu^{2} - \tau\nu^{2})} \right) + 
\nonumber
\\
&~~~
\frac{d\mu(\mu + \nu)}{\nu(\mu^{2} - \tau\nu^{2})} L_{k-1}^{1} \left( -\frac{ d^{2}(\mu + \nu)^{2}}{4\nu^{2}(\mu^{2} - \tau\nu^{2})} \right) 
\Bigg\rbrace
+ \frac{2\mu\nu\sqrt{\tau}}{\mu^{2} - \tau\nu^{2}} \left\lbrace L_{k} \left( -\frac{ d^{2}(\mu + \nu)^{2}}{4\nu^{2}(\mu^{2} - \tau\nu^{2})} \right) + L_{k-1}^{1} \left( -\frac{ d^{2}(\mu + \nu)^{2}}{4\nu^{2}(\mu^{2} - \tau\nu^{2})} \right) \right\rbrace +
\nonumber
\\
&~~~    \frac{ d^{2}(\mu + \nu)(\mu + \tau\nu)\sqrt{T}}{(\mu^{2} - \tau\nu^{2})^{2}} L_{k-1}^{1} \left( -\frac{ d^{2}(\mu + \nu)^{2}}{4\nu^{2}(\mu^{2} - \tau\nu^{2})} \right) 
+ \frac{ d^{2}\mu(\mu + \nu)^{2}\sqrt{\tau}}{\nu(\mu^{2} - \tau\nu^{2})} L_{k-2}^{2} \left( -\frac{ d^{2}(\mu + \nu)^{2}}{4\nu^{2}(\mu^{2} - \tau\nu^{2})} \right)
\Bigg]
\nonumber
\\
&= \frac{2\mu\nu\sqrt{\tau}}{\mu^{2} - \tau\nu^{2}} + \frac{2d^{2}\mu\nu(\mu + \tau\nu)^{2}\sqrt{\tau}}{(\mu^{4} - \tau^{2}\nu^{4})(\mu^{2} - \tau\nu^{2})}
+ \Bigg\lbrace \frac{2 d^{2}\mu^{2}(\mu + \nu)(\mu + \tau\nu)\sqrt{\tau}}{(\mu^{4} - \tau^{2}\nu^{4})(\mu^{2} - \tau\nu^{2})} + \frac{d^{2}(\mu + \nu)(\mu + \tau\nu)\sqrt{\tau}}{(\mu^{2} - \tau\nu^{2})^{2}} +
\nonumber
\\
&~~~~   \frac{ 2\mu\nu\sqrt{\tau}}{\mu^{2} - \tau\nu^{2}} \Bigg\rbrace
\frac{L_{k-1}^{1} \left( -\frac{ d^{2}(\mu + \nu)^{2}}{4\nu^{2}(\mu^{2} - \tau\nu^{2})} \right)}{L_{k} \left( -\frac{ d^{2}(\mu + \nu)^{2}}{4\nu^{2}(\mu^{2} - \tau\nu^{2})} \right)}
+ \frac{d^{2}\mu(\mu + \nu)^{2}\sqrt{\tau}}{\nu(\mu^{2} - \tau\nu^{2})^{2}}
\frac{L_{k-2}^{2} \left( -\frac{ d^{2}(\mu + \nu)^{2}}{4\nu^{2}(\mu^{2} - \tau\nu^{2})} \right)}{L_{k} \left( -\frac{ d^{2}(\mu + \nu)^{2}}{4\nu^{2}(\mu^{2} - \tau\nu^{2})} \right)}
\end{align}

\begin{align}
V_{C}^{x} &= \left\langle x_{1}x_{2} \right\rangle - \langle x_{1} \rangle \langle x_{2} \rangle
\nonumber
\\
&= \frac{2\mu\nu\sqrt{\tau}}{\mu^{2} - \tau\nu^{2}} 
+ \frac{ 2\mu\nu\sqrt{\tau}}{\mu^{2} - \tau\nu^{2}}  \frac{L_{k-1}^{1} \left( -\frac{ d^{2}(\mu + \nu)^{2}}{4\nu^{2}(\mu^{2} - \tau\nu^{2})} \right)}{L_{k} \left( -\frac{ d^{2}(\mu + \nu)^{2}}{4\nu^{2}(\mu^{2} - \tau\nu^{2})} \right)} +
\nonumber
\\
&~~~  \frac{d^{2}\mu(\mu + \nu)^{2}\sqrt{\tau}}{\nu(\mu^{2} - \tau\nu^{2})^{2}}
\left\lbrace  \frac{L_{k-2}^{2} \left( -\frac{ d^{2}(\mu + \nu)^{2}}{4\nu^{2}(\mu^{2} - \tau\nu^{2})} \right)}{L_{k} \left( -\frac{ d^{2}(\mu + \nu)^{2}}{4\nu^{2}(\mu^{2} - \tau\nu^{2})} \right)} 
-
\left( \frac{L_{k-1}^{1} \left( -\frac{ d^{2}(\mu + \nu)^{2}}{4\nu^{2}(\mu^{2} - \tau\nu^{2})} \right)}{L_{k} \left( -\frac{ d^{2}(\mu + \nu)^{2}}{4\nu^{2}(\mu^{2} - \tau\nu^{2})} \right)}  \right)^{2}
\right\rbrace
\end{align}

\begin{align}
V_{C}^{p} &= \left\langle p_{1}p_{2} \right\rangle = C_{i,j}^{m,n} \delta_{i,0}\delta_{j,1}\delta_{m,0}\delta_{n,1}
\nonumber
\\
&= \frac{1}{L_{k}\left( -\frac{d^{2}(\mu + \nu)^{2}}{4 \nu^{2} (\mu^{2} - \tau\nu^{2})} \right)}
\partial_{t} \left[ 
e^{ \frac{\mu^{2} + \tau\nu^{2}}{2(\mu^{2} - \tau\nu^{2})} t^{2}}
\partial_{b} \left[ 
 e^{ \frac{\mu^{2} + \tau\nu^{2}}{2(\mu^{2} - \tau\nu^{2})} b^{2} 
- \frac{2\mu\nu\sqrt{\tau}}{\mu^{2} - \tau\nu^{2}} b t} ~\times
 L_{k} \left( -\frac{ | d(\mu + \nu) + 2 i \nu (\nu\sqrt{\tau} t - \mu b) |^{2}}{4\nu^{2}(\mu^{2} - \tau\nu^{2})} \right)
\right]_{b=0}
\right]_{d=0}
\nonumber
\\
&= -\frac{2\mu\nu\sqrt{\tau}}{\mu^{2} - \tau\nu^{2}} 
- \frac{2\mu\nu\sqrt{\tau}}{\mu^{2} - \tau\nu^{2}} \frac{L_{k-1}^{1} \left( -\frac{ d^{2}(\mu + \nu)^{2})^{2}}{4\nu^{2}(\mu^{2} - \tau\nu^{2})} \right)}{L_{k} \left( -\frac{ d^{2}(\mu + \nu)^{2})^{2}}{4\nu^{2}(\mu^{2} - \tau\nu^{2})} \right)}
\end{align}
\end{subequations}
\end{widetext}

%%\bibliography{mdi}
\end{document}